\documentclass[11pt, a4paper, logo, copyright, nonumbering]{trillion}
\usepackage[authoryear, sort&compress, round]{natbib}
\usepackage{dblfloatfix}
\usepackage{ulem}
\usepackage{caption}
\usepackage{dramatist}
\usepackage{xspace}
\usepackage{pifont} 
\usepackage{booktabs}
\usepackage{multirow}
\usepackage{tcolorbox}
\usepackage{xltabular}
\usepackage{longtable}
\usepackage{hyperref}
\usepackage{amsfonts}
\usepackage{amsmath}
\usepackage{amssymb}
\usepackage{lineno}
\usepackage{multirow}
\usepackage{adjustbox}
\usepackage{graphicx}
\usepackage[bottom]{footmisc}

\usepackage{CJKutf8}
\usepackage{subfigure}
\usepackage{setspace}

\usepackage{dsfont}
\usepackage{array} 
\usepackage{tabularx} 
\usepackage{subfigure} 
\usepackage{xcolor} 

\usepackage{lipsum}  
\usepackage{multicol} 
\usepackage{xcolor}
\definecolor{mypurple}{HTML}{3C2C63}

\usepackage[T1]{fontenc}
\usepackage{amsmath,amsfonts,amssymb,amsthm,mathtools,bm}
\usepackage{booktabs}
\usepackage{nicefrac}
\usepackage{microtype}
\usepackage{xcolor}
\usepackage{enumitem}
\usepackage{multirow}
\usepackage{array}
\usepackage{color}
\usepackage[table]{xcolor} 
\usepackage{wrapfig}
\usepackage{multicol}
\usepackage{caption}
\usepackage{diagbox}
\usepackage{pifont}
\usepackage{pdfpages}
\usepackage{arydshln}
\usepackage{makecell}
\usepackage{graphicx}
\usepackage{adjustbox}
\usepackage{colortbl}
\usepackage{siunitx}
\usepackage{pifont}
\usepackage{seqsplit}
\usepackage{fancyvrb}
\tcbuselibrary{breakable}
\usepackage{fvextra}

\definecolor{blue(pigment)}{rgb}{0.2, 0.2, 0.6}
\definecolor{aliceblue}{rgb}{0.94, 0.97, 1.0}
\definecolor{lightgray}{rgb}{0.88, 0.88, 0.88}
\definecolor{piggypink}{rgb}{0.95, 0.9, 0.96}
\definecolor{mistyrose}{rgb}{1.0, 0.89, 0.88}
\definecolor{deeppurple}{rgb}{0.42, 0.13, 0.42}
\definecolor{lightgray}{rgb}{0.83, 0.83, 0.83}
\definecolor{pastelgray}{rgb}{0.81, 0.81, 0.77}
\definecolor{grey}{rgb}{0.5,0.5,0.5}
\definecolor{formatcolor}{rgb}{0.01, 0.31, 0.59}

\hypersetup{
  colorlinks=true,
  linkcolor=formatcolor,
  citecolor=formatcolor
}
\usepackage[capitalise,noabbrev,nameinlink]{cleveref}

\crefformat{figure}{Figure~#2{\color{formatcolor}#1}#3}
\crefformat{table}{Table~#2{\color{formatcolor}#1}#3}
\crefformat{equation}{Equation~#2{\color{formatcolor}#1}#3}
\crefformat{section}{Section~#2{\color{formatcolor}#1}#3}
\crefformat{appendix}{Appendix~#2{\color{formatcolor}#1}#3}

\newcommand{\huggingface}{\raisebox{-1.5pt}{\includegraphics[height=1.05em]{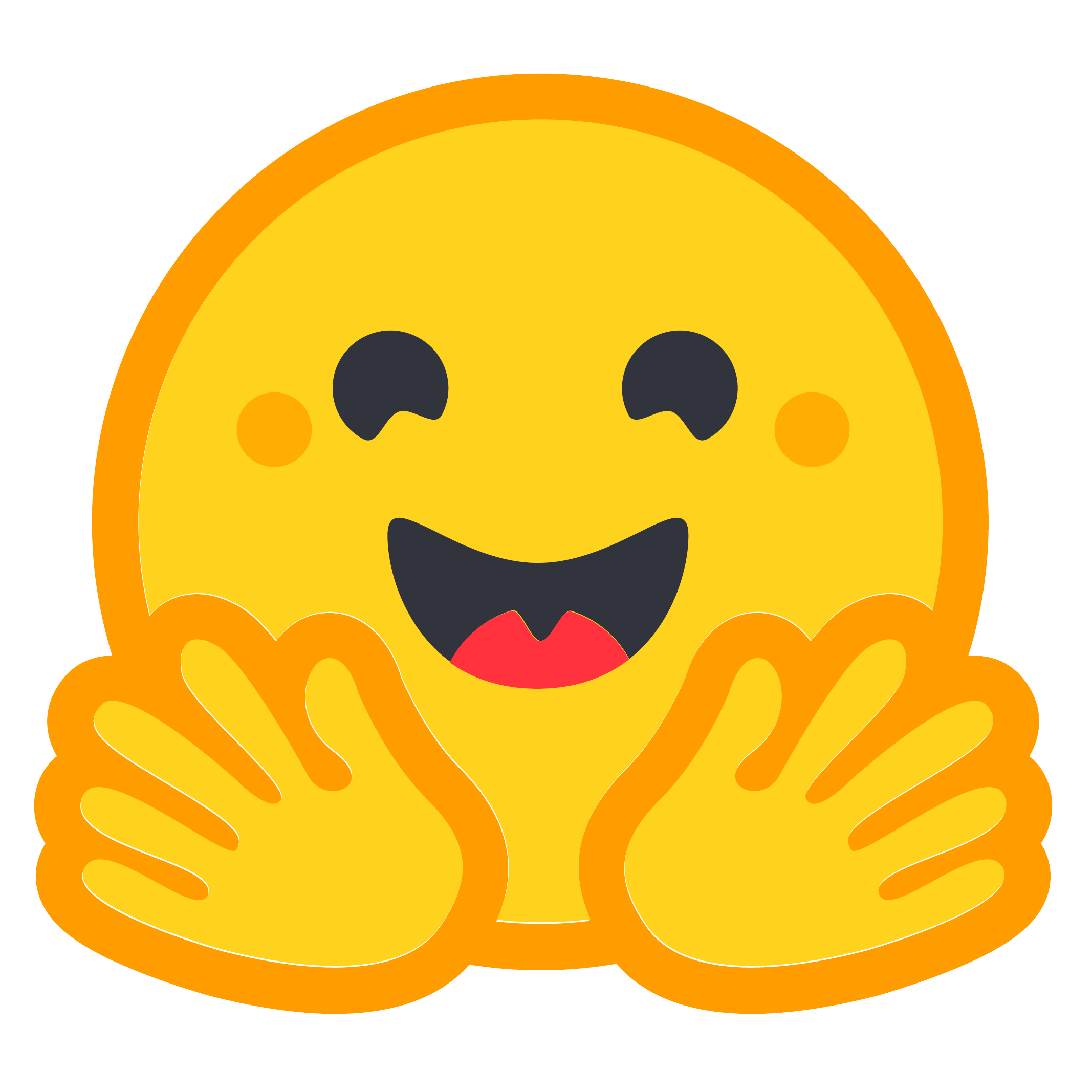}}\xspace}

\makeatletter
\def\@BTrule[#1]{%
  \ifx\longtable\undefined
    \let\@BTswitch\@BTnormal
  \else\ifx\hline\LT@hline
    \nobreak
    \let\@BTswitch\@BLTrule
  \else
     \let\@BTswitch\@BTnormal
  \fi\fi
  \global\@thisrulewidth=#1\relax
  \ifnum\@thisruleclass=\tw@\vskip\@aboverulesep\else
  \ifnum\@lastruleclass=\z@\vskip\@aboverulesep\else
  \ifnum\@lastruleclass=\@ne\vskip\doublerulesep\fi\fi\fi
  \@BTswitch}
\makeatother

\addto\extrasenglish{
    \def\sectionautorefname{Section}%

}

 {\begin{list}{}%
         {\setlength{\leftmargin}{#1}}%
         \item[]%
 }
 {\end{list}}
 
\reportnumber{001} 

\renewcommand{\today}{}

\title{\centering \textsc{TheBioCollection}: \\ Unified Pre-Training Scale LLM Corpus for Biology}

\author[*]{
Hyunjin Seo$^{1,*}$,
Hyeon Hwang$^{1,*}$,
Gyubok Lee$^{2}$,
Jay Shin$^{1}$,
Jimin Park$^{3}$,
Taesoo Kim$^{4}$,
Sanghoon Lee$^{5}$,
Hongjoon Ahn$^{1,\ddagger}$,
Sungjun Han$^{1,\ddagger}$,
Sangwon Jung$^{1,\ddagger}$
\\
\small
{$^{1}$ Trillion Labs}
{$^{2}$ KAIST}
{$^{3}$ SK Biopharmaceuticals Co., Ltd.}
{$^{4}$ Lunit Inc.}
{$^{5}$ AIGEN Sciences Inc.}
\\
\texttt{\small{$^{*}$ First Author}} 
\\
\texttt{\small{$^{\ddagger}$ Project Lead}}
}

\date{}

\begin{abstract}

The push toward large language models for biology (BioLM) has created a need for training corpora that can endow models with a genuine understanding of biology. However, existing biological resources, such as molecular databases, protein repositories, genomic annotations, single-cell atlases, and pathway databases, are scattered across heterogeneous formats and remain unorganized into a cohesive corpus for language model training. We present \system{}, a 52.6B-token pre-training-scale corpus that converts these disparate resources into a unified, training-ready form spanning small molecules, proteins, genomic sequences, cells, and pathways. Beyond consolidating existing data, \system{} enriches each record with tool-computed biological properties and introduces new instruction tasks for capabilities that current corpora barely cover. We pair the corpus with \evalsuite{}, a matched suite probing recognition, generation, and prediction across molecular, protein, genomic, cellular, and cross-domain settings. Holding the base Gravity-16B-A3B architecture fixed, training on \system{} more than doubles its overall score on \evalsuite{} with gains in every domain, while leaving general linguistic ability nearly intact.

\noindent\textbf{\huggingface Training corpus:}
\href{https://huggingface.co/datasets/trillionlabs/TheBioCollection}
{\textcolor{mypurple}{\system{}}}

\noindent\textbf{\huggingface Evaluation dataset:}
\href{https://huggingface.co/datasets/trillionlabs/TheBioCollection-Eval}
{\textcolor{mypurple}{\evalsuite{}}}

\noindent\textbf{\huggingface Model:}
\href{https://huggingface.co/trillionlabs/Gravity-bio-16B-A3B}
{\textcolor{mypurple}{Gravity-bio-16B-A3B}}
\end{abstract}

\begin{document}
\begin{CJK*}{UTF8}{mj}

\maketitle


\section{Introduction}\label{1_intro}

Large language models (LLMs) for biology (BioLM)~\citep{naturelm, scireasoner, txgemma, jang2026towards, logos} aim to extend LLMs beyond general text understanding toward the representation and reasoning of biological systems and generation of biological entities. Unlike task-specific predictors trained for a single modality or endpoint in biology~\citep{esm2,scgpt, dnabert2}, BioLM seek to operate over diverse biological entities and processes, including small molecules, proteins, genomic sequences, cell states, and pathways. Understanding this broader spectrum is important as many biological questions involve not only recognizing isolated entities, but also understanding their properties, functions, interactions, and effects across different biological contexts~\citep{primer, mazein2024graph, mohamed2021biological}.

Realizing this vision requires a large-scale training corpus for biology that exposes the model to the breadth and structure of biology. The field has already produced many useful resources, including molecular databases, protein repositories, genomic annotations, single-cell atlases, pathway databases, and computational biology tools. 
However, these resources are rarely organized as a large-scale, LLM training-friendly data corpus for BioLM; instead, they are scattered across heterogeneous formats such as tables, sequence files, graph edges, structured annotations, or isolated instruction schemas~\citep{smolinstruct, mollangbench, vibeproteinbench, proteinlmbench, chatnt, tabulasapiens}. 
Nevertheless, no prior effort has consolidated these resources into a large-scale pre-training corpus that teaches BioLMs broad spectrum of biological knowledge spanning molecules, proteins, genomes, cells, and their cross-domain relationships.

To address this gap, we introduce \system{}, a 52.6B-token pre-training scale corpus for biology that turns heterogeneous biological resources into LLM training friendly data. \system{} is built through a construction pipeline that collects resources across biological domains, refines them through deduplication, entity tagging, and entity augmentation, enriches them with tool-computed biological properties, and renders them as instruction-form data with programmatically verifiable answers. 
This pipeline makes the resulting corpus both broad in biological coverage and directly usable for training BioLMs.
To systematically assess biological capability after training, we also release \evalsuite{}, a separate evaluation suite matched to the breadth of \system{}. The suite spans recognition, generation, and prediction across molecular, protein, genomic, cellular, and cross-domain settings. 


\begin{figure}[t!]
    \centering
    \includegraphics[width=1.0\linewidth]{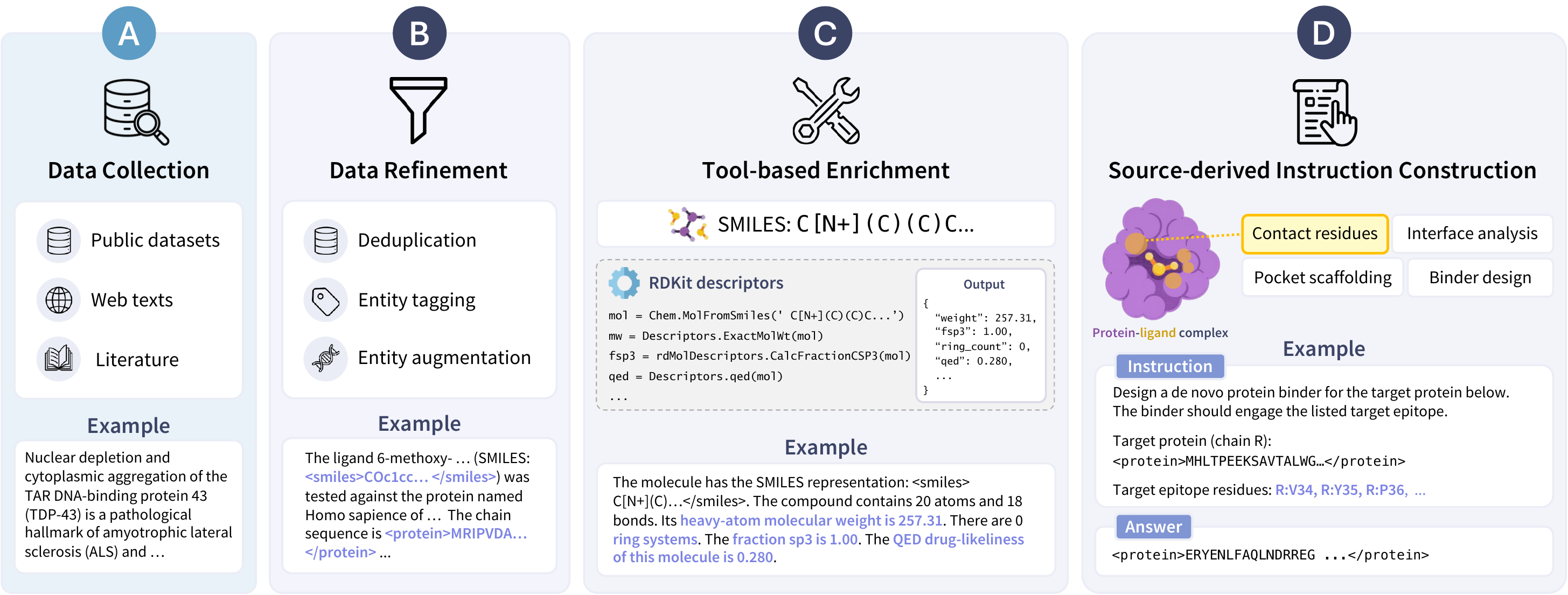}
    \caption{The construction pipeline of \system{}. We first integrate scattered biological resources and refine them into structured texts. Subsequently, we enrich the biological information in free-text stream with computational tools. Finally, we construct new instruction datasets for domains that are underrepresented in the corpora.}
    \label{fig:dataset_construction}
\end{figure}
To isolate the effect of the corpus, we hold the base architecture of Gravity-16B-A3B~\citep{gravity-moe-2026} fixed and train it on \system{}, comparing the result against the base model. 
Our evaluations using \evalsuite{} reveal that the model trained with our corpus improves  task performance on biology without specialized architectures or additional training tricks. Specifically, training on \system{} more than doubles the base model's overall score with consistent gains in every biological domain. Furthermore, in our ablation study, the model trained with our corpus outperforms the model further trained only with web-text on biological capability while preserving general linguistic ability with minimal degradation. These results suggest that biological capability and general linguistic ability are not mutually exclusive and can be learned jointly without interference.

To summarize, our contributions are as follows:
\begin{itemize}[topsep=0pt,itemsep=1mm, parsep=0pt, leftmargin=5mm]
    \item \textbf{\system{} corpus.} We introduce \system{}, a  52.6B-token biological corpus that turns scattered public datasets into LLM-ready training data, spanning small molecules, proteins, genomic sequences, cells, and pathways, with tool-computed biological properties exposed directly in texts. We further construct new instruction datasets for tasks that existing corpora cover poorly, including protein binding and genomic feature localization.
    \item \textbf{\evalsuite{}.} We release \evalsuite{}, a matched evaluation suite that covers recognition, generation, and prediction across molecular, protein, genomic, cellular, and cross-domain settings, covering a broad spectrum of biological applications.
    \item \textbf{Controlled evidence of corpus effect.} Holding the base LLM architecture fixed, we show that training on \system{} improves broad biological task performance without additional training tricks, while maintaining general linguistic performance through the inclusion of only a limited amount of web text in the training corpus.
\end{itemize}

\begin{figure}[t!]
    \centering
    \includegraphics[width=\linewidth]{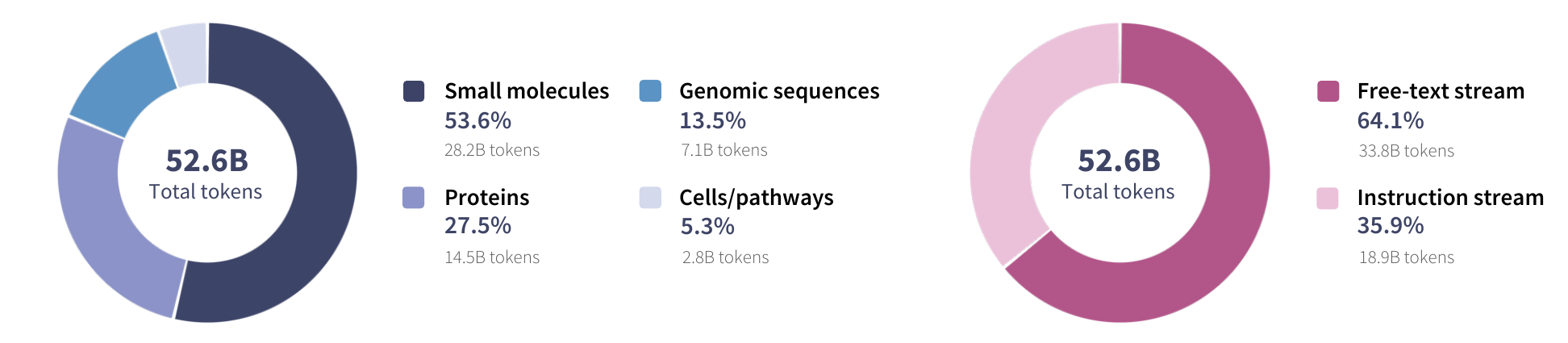}
    \caption{Token statistics of \system{}. \textit{Left}: the 52.6B-token corpus split across small molecules, proteins, genomic sequences, and cells and pathways. \textit{Right}: the same corpus split into free-text and instruction streams. Token counts exclude scientific literature used during training. }
    \label{fig:dataset_stats}
\end{figure}

\section{Dataset Construction}\label{2_method}


\system{} is a 52.6B-token corpus that places small molecules, proteins, genomic sequences, cells, pathways, and cross-domain relationships under a shared language interface. Its composition is summarized in \cref{fig:dataset_stats}. We build the corpus by collecting public, commercially usable resources across all five domains (\cref{2.1}) and transforming them into LLM training friendly data through three complementary steps: refining database records into self-contained descriptions (\cref{2.2}), enriching them with tool-computed biological features (\cref{2.3}), and constructing instruction datasets from both existing resources and newly generated datasets that no existing instruction-tuning corpus covers (\cref{2.4}). To measure whether these data translates into biological understanding, we pair the corpus with \evalsuite{}, a matched evaluation suite spanning the same domains together with cross-domain reasoning (\cref{2.5}).


\begin{table}[t]
\centering
\footnotesize
\caption{External resources used to construct free-text stream in \system{}, grouped by biological domain.}
\label{tab:source_list}
\setlength{\tabcolsep}{4pt}
\renewcommand{\arraystretch}{1.15}
\begin{tabularx}{\linewidth}{
  >{\raggedright\arraybackslash\bfseries}p{0.27\linewidth}
  >{\raggedright\arraybackslash}X
}
\toprule
Domain & \textbf{Source datasets} \\
\midrule
Small molecules &
PubChem \citep{pubchem}; DrugBank \citep{drugbank}; BindingDB~\citep{bindingdb}; ChEMBL~\citep{chembl}; USPTO-50K~\citep{uspto50k}.\\
\midrule
Proteins &
UniProt Knowledgebase \citep{uniprot}; AlphaFoldDB~\citep{afdb}; Protein Data Bank (PDB)~\citep{pdb}. \\
\midrule
Genomic sequences &
GENCODE \citep{gencode}; ENCODE \citep{encode}; RNAcentral \citep{rnacentral}; Rfam \citep{rfam}. \\
\midrule
Cells/Pathways &
L1000~\citep{lincs_l1000};
Human Protein Atlas (HPA) single-cell type data~\citep{humanproteinatlas};
GeneRIF~\citep{generif};
NCBI Gene~\citep{ncbi_gene};
Gene Ontology / GOA human~\citep{go};
Cell Ontology~\citep{cell_ontology}; CELLxGENE papers \citep{cellxgene}; JUMP Cell Painting~\citep{jumpcellpainting}; HuBMAP~\citep{hubmap}.\\
\midrule
Cross-domain KGs &
Hetionet \citep{hetionet}; DRKG \citep{drkg}; DGIdb \citep{dgidb};
STRING \citep{string}; OmniPath \citep{omnipath}; Reactome \citep{reactome}. \\
\midrule
Broad scientific literature &
PubMed; bioRxiv; medRxiv. \\
\bottomrule
\end{tabularx}
\end{table}

\subsection{Data  collection}\label{2.1}

We build \system{} only from public resources that permit commercial use, and we keep each entity's source identifiers throughout so every record stays traceable to its origin. The collected resources span five domains targeted by the corpus, including small molecules, proteins, genomic sequences, cells/pathways, and cross-domain relationships. As summarized in \cref{tab:source_list}, these sources provide complementary biological signals, ranging from molecular characteristics to protein sequences and functions, genomic and transcriptomic annotations, cellular identities and perturbation responses, and knowledge-graph-linked relationships across biological entities. Finally, to retain general scientific and language ability that structured data alone does not provide, we include broad scientific literature from PubMed\footnote{\url{https://pubmed.ncbi.nlm.nih.gov/}}, bioRxiv\footnote{\url{https://www.biorxiv.org/}}, and medRxiv\footnote{\url{https://www.medrxiv.org/}}. These scientific sources are included for training but are excluded from the token counts reported for \system{}. The remaining subsections describe how we transform these resources into LLM training friendly data and enrich them with additional biological signals beyond public resources.




\subsection{Dataset refinement}\label{2.2}

Curated biological databases contain rich biological knowledge, but often in formats that language models cannot consume directly~\citep{lincs_l1000, pubchem, bindingdb, uspto50k, chembl}. We therefore transform each database record into a self-contained natural-language description that consolidates entity's key attributes into a single textual record. Depending on the modality, these descriptions may include physicochemical properties, perturbation profiles, structural information, or functional annotations. We then render each value with a plain language statement of what it indicates. Symbolic identifiers including SMILES strings and protein, DNA, or RNA sequences are placed inline and wrapped in dedicated tagging tokens (\textit{e.g.}, \texttt{<smiles>...</smiles>}) that keep them distinct from the surrounding text, following~\cite{naturelm, scireasoner}. The resulting records keep an entity's sequence, properties, and interpretation within a single context, turning structured database entries into language the model can learn from.

Beyond single-entity descriptions, we also link entities across domains by gathering evidence about the same entity from several databases into one record. A knowledge graph serves as the skeleton and the source databases as the payload: we use cross-domain knowledge graphs (Hetionet~\citep{hetionet}, DRKG~\citep{drkg}, STRING~\citep{string}, OmniPath~\citep{omnipath}) to decide which entities belong together through typed relations such as compound--target binding, gene--pathway participation, protein--protein interaction, and transcription-factor--target regulation, then fill in each linked entity with the records already held in the underlying databases. As a result, a single record reads as a mechanistic chain rather than an isolated fact: a protein anchor, for instance, carries its function and pathway membership, the compounds that bind it with their measured affinities, and the transcriptional response those perturbations induce in a given cell line, so that molecule, protein, pathway, and cell-domain evidence appear together in one context window.

\subsection{Tool-computed feature narratives}\label{2.3}

We verbalize biological signals that live in databases and computational tools rather than in text. We run standard computational-biology tools over source records and render their outputs as natural-language narratives with explicit provenance, exposing structured, tool-computable signals that public free text rarely contains. We apply the same recipe across modalities; the full feature sets and extraction pipelines are given in \cref{B}.

\paragraph{Small molecules.} We enrich the refined PubChem molecular narratives produced from \cref{2.2} with RDKit-computed descriptors~\citep{rdkit}.  After parsing and canonicalization, we derive deterministic features such as ring-system composition, scaffold and complexity measures, stereochemistry, functional-group and structural-alert matches, and fingerprint summaries, and incorporate them into the corresponding narratives.

\paragraph{Proteins.} We construct protein structure/function narratives by linking UniProt sequences to AlphaFold database and Protein Data Bank (PDB) structures. These records combine deterministic structure-derived features such as physicochemical properties, secondary structure, fold topology, and solvent exposure using Biopython~\citep{biopython} and DSSP~\citep{dssp}, with UniProt Knowledgebase-derived annotations such as domains, InterPro/Pfam families, transmembrane regions, GO terms, and subcellular localization~\citep{uniprot}.

\paragraph{Genomic sequences.}
We represent regulatory DNA intervals and RNA transcripts as structured biological entities. For DNA, we use Python-based processing pipelines built on indexed FASTA access (pyfaidx~\citep{pyfaidx}), Biopython parsing, and deterministic GTF/BED interval handling to associate genomic sequences from GENCODE~\citep{gencode} and ENCODE~\citep{encode} with coordinate and assembly metadata, regulatory evidence, and sequence-derived features such as GC content, CpG statistics, entropy, homopolymer length, and genomic feature overlap. For RNA, we build transcript-level narratives from GENCODE and RNA resources including RNAcentral~\citep{rnacentral} and Rfam~\citep{rfam}. These narratives incorporate features derived from Biopython-based FASTA and GenBank parsing~\citep{genbank}, codon and ORF analyses, and sequence-motif profiling.

\paragraph{Cellular profiles.}
We convert cellular profiles into language that captures four signals: expression, perturbation response, spatial context, and morphology. Single-cell records describe cell states from marker genes and tissue context~\citep{humanproteinatlas, cell_ontology}. Perturbation records link CRISPRa/CRISPRi interventions to the differential-expression signatures they induce~\citep{perturbench,norman,replogle}. Spatial records from HuBMAP~\citep{hubmap} place each measured spot in context, reporting its tissue, coordinates, local neighborhood, and spot-local transcript features. We build them from AnnData/H5AD inputs~\citep{anndata} and recover each neighborhood with SciPy KD-tree radius queries~\citep{scipy}. JUMP Cell Painting records~\citep{jumpcellpainting} capture how each perturbation shifts cell morphology. We derive them from CellProfiler profiles~\citep{cellprofiler}, normalize features against controls with pandas/NumPy~\citep{pandas,numpy}, and retrieve similar morphologies with scikit-learn nearest-neighbor search~\citep{sklearn}. 


\begin{table}[t]
\centering
\footnotesize
\caption{Imported instruction datasets curated for \system{}.}
\label{tab:instruction_sources}
\setlength{\tabcolsep}{5pt}
\renewcommand{\arraystretch}{1.18}
\begin{tabularx}{\linewidth}{
  >{\raggedright\arraybackslash\bfseries}p{0.17\linewidth}
  >{\raggedright\arraybackslash}p{0.4\linewidth}
  >{\raggedright\arraybackslash}X
}
\toprule
Domain & \textbf{Source datasets} & \textbf{Task coverage} \\
\midrule
Small molecules &
Mol-Instructions~\citep{Mol-instructions}; SMolInstruct~\citep{smolinstruct};
MolLangBench~\citep{mollangbench}; ChEBI-20-MM~\citep{chebi20};
ChemData700K~\citep{chemdata700k}; Language-Plus-Molecules (LPM-24)~\citep{lpm24};
TxGemma~\citep{txgemma} &
Text-guided molecule design and generation, molecule captioning, SMILES/name/text conversion,
atom- and functional-group recognition, molecular property and toxicity prediction,
forward synthesis, retrosynthesis, and reaction prediction. \\
\midrule
Proteins &
ProteinLMBench~\citep{proteinlmbench}; BioReason-Pro~\citep{bioreasonpro};
TxGemma~\citep{txgemma} &
Protein sequence-to-function prediction, protein function description, and drug--target or affinity-style prediction. \\
\midrule
Cells/pathways & Cell2Sentence~\citep{cell2sentence}; 
ENCODE-C2S~\citep{encode}; Tabula Sapiens~\citep{tabulasapiens};
PerturBench Norman and Replogle K562~\citep{perturbench,norman,replogle};
TxGemma~\citep{txgemma} &
Cell-type classification, tissue-aware cell-state interpretation, perturbation-response and differential-expression signature prediction, drug synergy prediction, and omics response modeling. \\
\bottomrule
\end{tabularx}
\end{table}
\subsection{Instruction-tuning datasets}\label{2.4}

The instruction corpus draws on two complementary sources. The first is imported instruction datasets: the public datasets refined via \cref{2.2}, which already expose biological entities through natural-language instructions. The second is source-derived instruction tasks that we construct ourselves. Across the full instruction corpus, approximately 30\% of examples include few-shot support and the rest are zero-shot, following the FLAN collection~\citep{flan_collection}.

For the imported datasets, our guiding principle is careful curation. We select commercially usable datasets that broaden the task coverage of \system{} and add them to the corpus with associated tagging tokens. \cref{tab:instruction_sources} summarizes these datasets and the tasks they cover. The source-derived tasks, by contrast, are generated directly from non-text biological resources, and we design them so that answers stay structured, operational, and programmatically checkable. Crucially, these tasks cover capabilities that existing public biological instruction datasets rarely address, broadening the task coverage of \system{}. We detail two such task families below, and \cref{C} gives full construction details.

\paragraph{Protein binding.}
We derive binding tasks from protein--ligand, protein--protein, and protein--peptide complexes by computing target--binder interfaces from heavy-atom contacts. The resulting instructions ask the model to recover masked binding-site residues, summarize contact patterns, scaffold binding pockets, or generate binder sequences conditioned on target epitope residues. After structural-quality filtering, deduplication, and evaluation-overlap removal, this yields 12M binding instruction records before context-length filtering.


\paragraph{DNA/RNA feature localization.}
We cast regulatory DNA and transcript RNA annotations as exact span-recovery tasks. Given a sequence window, the model returns the feature label, coordinates, and subsequence, making every answer programmatically checkable. DNA tasks cover regulatory elements, open-chromatin peaks, splice-related sites, and transcription-factor motifs; RNA tasks cover coding and untranslated regions, exon junctions, family hits, miRNAs, and tRNA anticodons. We discard ambiguous or low-quality windows and any record that fails deterministic span self-checks.






\subsection{Evaluation suite: \evalsuite{}}\label{2.5}
\begin{figure}[t!]
    \centering
    \includegraphics[width=1.0\linewidth]{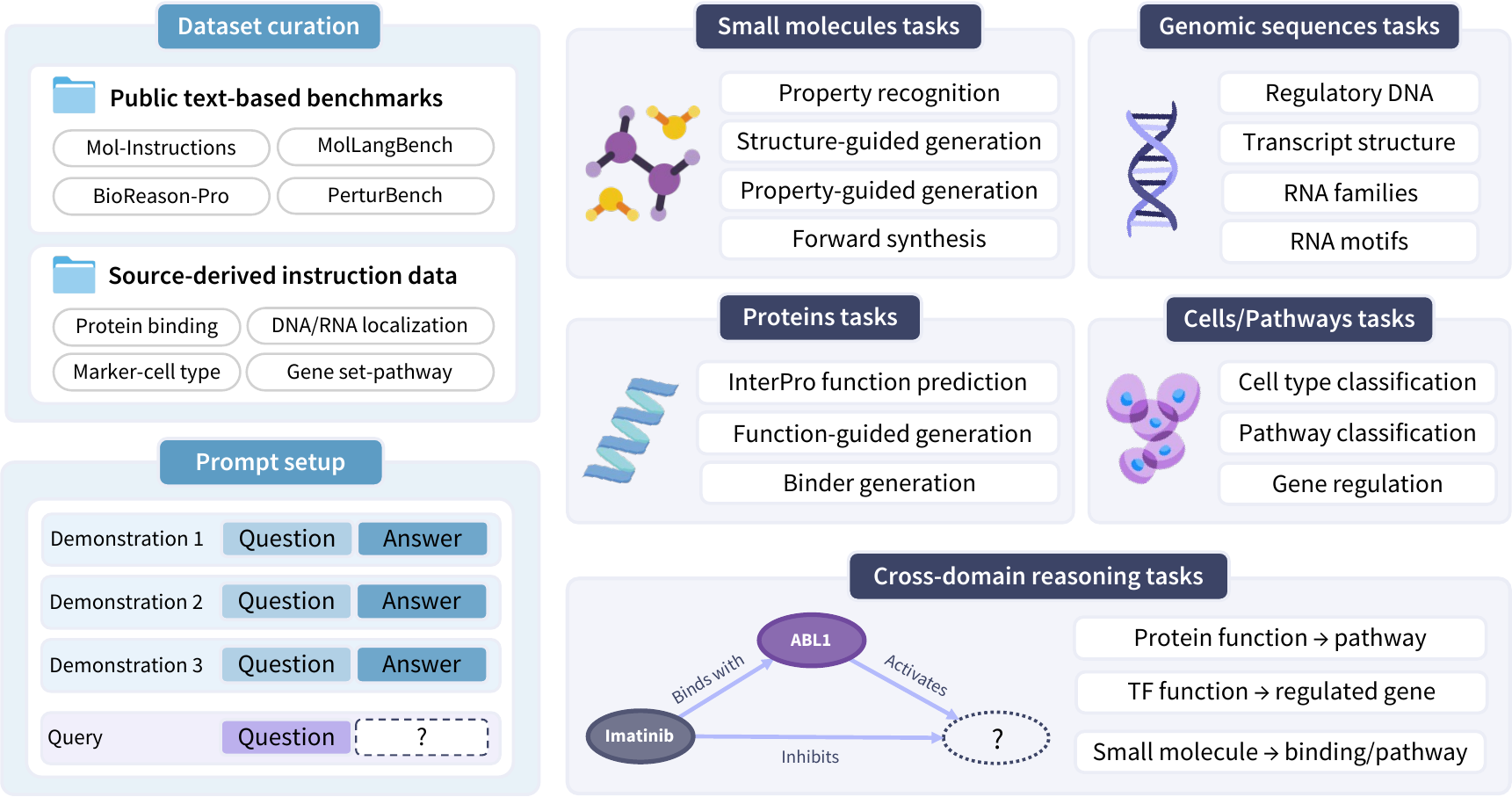}
    \caption{Illustration of \evalsuite{}. \textit{Left}: Evaluation dataset construction. We combine established public benchmarks with the instruction datasets that we newly construct from primary biological sources, and convert single-domain tasks into a 3-shot format; cross-domain tasks are evaluated zero-shot. \textit{Right}: Evaluation task organization. Within each domain, we adopt tasks that are commonly studied in the corresponding domain-specific literature. To probe whether models transfer knowledge across domains, we further introduce cross-domain reasoning tasks constructed from biological knowledge graphs.}
    \label{fig:eval_construction}
\end{figure}

\evalsuite{} measures whether training on \system{} improves biological reasoning capability across domains of organization. We build it by drawing the subtasks from many scattered existing benchmarks and combining them with tasks we construct ourselves, then standardizing everything into one cross-domain suite. The construction of evaluation datasets and the task organization are illustrated in \cref{fig:eval_construction}. This consolidates evaluations that were previously fragmented across separate resources and formats, and it mirrors the breadth of the corpus, spanning small molecules, proteins, genomics, and cells together with cross-domain reasoning. We sample 100 and 50 records per subtask in single-domain and cross-domain, respectively, yielding 1,650 examples across 18 tasks, and for any task whose format also appears in training, we separate train and test entities to prevent overlap. The statistics of the evaluation dataset are shown in \cref{tab:eval-statistics}. We show an example question for each subtask in \cref{tab:eval-question-examples}, and provide per-task evaluation metrics in \cref{E}. We group the tasks by biological domain as follows:

\begin{itemize}[topsep=0pt,itemsep=1mm, parsep=0pt, leftmargin=5mm]
\item \textbf{Small molecules:} The small-molecule tasks span recognition, generation, and reaction prediction. They cover molecular recognition and reconstruction from structural descriptions~\citep{mollangbench}, description-guided molecule design~\citep{Mol-instructions}, and forward synthesis~\citep{smolinstruct}.

\item \textbf{Proteins:} The protein tasks evaluate function prediction and design. Function prediction asks the model to predict the natural-language names of a protein's InterPro IDs from its sequence. Design tasks consist of functional description-conditioned generation~\citep{Mol-instructions} and binder generation against small-molecule and protein targets.

\item \textbf{Genomic sequences:} The genomics tasks evaluate DNA and RNA feature annotation as structured span recovery. Specifically, the DNA tasks cover candidate cis-regulatory element (cCRE), open-chromatin, and splice-site localization, whereas the RNA tasks require identifying Rfam families and tRNA anticodons. Each task asks the model to identify the biological feature type and return its label, exact coordinates, and corresponding subsequence, mirroring the interface used during training.

\item \textbf{Cells and pathways:} The cell and pathway tasks evaluate marker-based cell-type identity~\citep{sctype}, Hallmark pathway recognition~\citep{msigdb_hallmark,enrichr}, and K562 CRISPRi perturbation-response direction~\citep{replogle}. Each is posed as multiple-choice or as up/down classification over candidate genes.

\item \textbf{Cross-domain reasoning:} The cross-domain tasks test whether the model can chain evidence across domains rather than within a single one. Each is a two-hop multiple-choice question linking an entity's function or identity to a knowledge-graph relation: a protein's function to a pathway it participates in, a transcription factor's function to a gene it regulates, and a small molecule to the target and pathway it acts through~\citep{drkg,reactome,omnipath}.

\end{itemize}


\paragraph{Filtering and decontamination.} 
We clean the corpus in three steps:
\begin{enumerate}[topsep=2pt,itemsep=1mm,parsep=0pt,leftmargin=*,label=(\arabic*)]
\item \textbf{Validity checks.} Every record passes source-specific checks: malformed molecules and protein sequences are dropped, binding records must satisfy contact and interface constraints, and a DNA/RNA span is kept only when its target feature is unambiguously recoverable.
\item \textbf{Deduplication.} Reading the corpus one record at a time, we assign each a stable key (normalized text hash, canonical molecule identifier, sequence hash, genomic coordinate, document ID, or instruction--answer hash) and drop any record whose key has already appeared, along with residual leakage and artifact rows.
\item \textbf{Decontamination.} We hold \evalsuite{} out of training with task-specific keys and exact string matching, e.g., genomics by sequence hash, accession, and coordinate, and protein binder-generation targets by excluding any exact, subsequence, or shared 15-mer overlap.
\end{enumerate}

\begin{table*}[t]
\centering
\footnotesize
\caption{Performance across diverse biological domains: \textbf{Base} vs.\ \textbf{Ours}. \textbf{Base} is the base model checkpoint before training; \textbf{Ours} adds the \system{} corpus on top of general scientific and web text. Scores are averaged first across metrics within a task, then across subtasks within a domain; the overall score averages all subtasks. Full subtask-by-metric results are provided in \cref{D}. Bold indicates the best model in each row.}
\label{tab:bmfm-main-results-vanilla-vs-ours}

\setlength{\tabcolsep}{5pt}
\renewcommand{\arraystretch}{0.95}

\begin{tabularx}{0.92\textwidth}{
>{\raggedright\arraybackslash\bfseries}p{0.2\textwidth}
>{\raggedright\arraybackslash}X
r
r
r
}
\toprule
Domain & \textbf{Task} & \textbf{Base} & \textbf{Ours} & $\Delta$ \\
\midrule

\multirow{4}{*}{Small molecules}
& Molecule reconstruction/design & 0.200 & \textbf{0.522} & +0.322 \\
& Forward synthesis & 0.213 & \textbf{0.619} & +0.406 \\
& Molecular property recognition & 0.280 & \textbf{0.390} & +0.110 \\
\cmidrule(lr){2-5}
& \textit{Domain average} & 0.223 & \textbf{0.513} & +0.290 \\
\midrule

\multirow{4}{*}{Proteins}
& Text-conditioned functional protein design & 0.243 & \textbf{0.522} & +0.279 \\
& Binder design & 0.234 & \textbf{0.645} & +0.411 \\
& Protein function prediction & 0.000 & \textbf{0.055} & +0.055 \\
\cmidrule(lr){2-5}
& \textit{Domain average} & 0.159 & \textbf{0.407} & +0.248 \\
\midrule

\multirow{3}{*}{Genomic sequences}
& DNA regulatory/splice span localization & 0.134 & \textbf{0.516} & +0.382 \\
& RNA family/anticodon span localization & 0.238 & \textbf{0.396} & +0.158 \\
\cmidrule(lr){2-5}
& \textit{Domain average} & 0.175 & \textbf{0.468} & +0.293 \\
\midrule

\multirow{4}{*}{Cells/pathways}
& Cell type recognition & 0.470 & \textbf{0.580} & +0.110 \\
& Hallmark program recognition & 0.520 & \textbf{0.750} & +0.230 \\
& Perturbation response prediction & 0.015 & \textbf{0.498} & +0.483 \\
\cmidrule(lr){2-5}
& \textit{Domain average} & 0.335 & \textbf{0.609} & +0.274 \\
\midrule

\rowcolor{aliceblue}
\textbf{Overall}
& 
& 0.223
& \textbf{0.499}
& +0.276 \\
\bottomrule
\end{tabularx}
\end{table*}
\section{Experiments}

\subsection{Training details}

We train the base Gravity-16B-A3B~\citep{gravity-moe-2026} on \system{}. We choose Gravity-16B-A3B for two reasons: it exposes a pre-anneal checkpoint, and its pretraining is known to contain no biological corpus, so any biological capability we observe comes from our data rather than the base. Beginning before annealing matters as most high-impact learning happens during annealing, so starting before it lets us measure the annealing effect of \system{} cleanly without the effect of another corpus entangling with ours. We mix our corpus with general scientific and web text from OLMoCR science PDFs, DCLM-CC, EAI-DCLM, PubMed, bioRxiv, and medRxiv as replay, to counteract forgetting and preserve broad language ability. We use a sequence length of 8{,}192 and a global batch of 8.4M tokens ($1{,}024 \times 8{,}192$), an annealed learning rate of $4 \times 10^{-4}$, and weight decay of 0.01, trained on 16 NVIDIA B200 GPUs.

To attribute the measured gains to the corpus rather than to the training process itself, we compare against the base model under the same architecture. \textbf{Base} is the base Gravity-16B-A3B checkpoint without any training, and \textbf{Ours} is the model trained on \system{}, starting from that same checkpoint.

\subsection{Evaluation protocol} 

For all evaluation tasks except cross-domain reasoning, we adopt a fixed three-shot prompting setting, prepending three task-specific demonstration examples to each query; the cross-domain tasks are evaluated zero-shot. We then aggregate by averaging within a subtask, across subtasks within a domain, and finally across all subtasks for the overall score. To keep these aggregated scores directionally consistent, we average only higher-is-better metrics and exclude the one lower-is-better metric, SMILES Levenshtein distance, which we still report per subtask in \cref{D}.

\subsection{Main results}
\label{sec:main_results}

We report the performance of the model trained with our corpus in \cref{tab:bmfm-main-results-vanilla-vs-ours}. As shown, training on \system{} improves performance at every biological domain. The model raises the overall score from 0.223 to 0.449, more than doubling the baseline, with consistent domain gains across genomic sequence (0.175 $\rightarrow$ 0.468), small molecule (0.223 $\rightarrow$ 0.513), protein (0.159 $\rightarrow$ 0.407), cell/pathway (0.335 $\rightarrow$ 0.609).

The gains concentrate on tasks whose supervision is structured and tool-derived rather than literature-style text. This is especially pronounced in DNA regulatory/splice-site localization (0.134 $\rightarrow$ 0.516) and protein binder design (0.234 $\rightarrow$ 0.645), where the supervision is difficult to retrieve from public free-text records and is instead supplied by matched instruction tasks. Beyond these, the corpus also improves recognition tasks such as molecular property recognition (0.280 $\rightarrow$ 0.390) and cell-type and hallmark-program recognition (0.470 $\rightarrow$ 0.580, 0.520 $\rightarrow$ 0.750), for which the tool-driven narratives (\cref{2.2}) supply the signal implicitly: RDKit descriptors render each molecule's computed structural properties into text, and marker-gene narratives describe each cell by its expression state. Although these narratives are written as descriptions rather than task-formatted supervision, they teach the property and identity cues that recognition requires.


\subsection{Ablation study}

\paragraph{Contribution of the sole biological corpus.}
\begin{wraptable}{r}{0.4\linewidth}
\vspace{-\baselineskip}
\centering
\footnotesize
\setlength{\tabcolsep}{5pt}
\renewcommand{\arraystretch}{1.0}
\caption{Scientific free text used in annealing.}
\label{tab:scientific-text}
\begin{tabular}{lrr}
\toprule
\textbf{Source} & \textbf{Docs} & \textbf{Tokens} \\
\midrule
PubMed  & 6.43M & 48.8B \\
bioRxiv & 0.47M & 4.6B \\
medRxiv & 0.11M & 0.58B \\
\midrule
\textbf{Total} & \textbf{7.01M} & \textbf{54.0B} \\
\bottomrule
\end{tabular}
\vspace{-\baselineskip}
\end{wraptable}

\begin{table*}[t]
\centering
\footnotesize
\caption{Performance comparison: \textbf{Text-annealing only (TA only)} vs.\ \textbf{Ours}. \textbf{TA only} is trained on general scientific and web text alone, with no \system{} data; \textbf{Ours} adds the \system{} corpus on top of that same general text. Since the two share the same base checkpoint, annealing process, and general text, the gap between them isolates the contribution of the \system{} corpus. For biological domains, scores are averaged first across metrics within a task, then across subtasks within a domain; the overall score averages all subtasks. General-language benchmarks are reported on their native scales and excluded from the biological overall score. Bold indicates the best model in each row.}
\label{tab:textanneal-vs-ours}
\setlength{\tabcolsep}{5pt}
\renewcommand{\arraystretch}{0.95}
\begin{tabularx}{0.92\textwidth}{
>{\raggedright\arraybackslash\bfseries}p{0.2\textwidth}
>{\raggedright\arraybackslash}X
r
r
r
}
\toprule
Domain & \textbf{Task} & \textbf{TA only} & \textbf{Ours} & $\Delta$ \\
\midrule
\multirow{4}{*}{Small molecules}
& Molecule reconstruction/design & 0.432 & \textbf{0.522} & +0.090 \\
& Forward synthesis & 0.477 & \textbf{0.619} & +0.142 \\
& Molecular property recognition & 0.300 & \textbf{0.390} & +0.090 \\
\cmidrule(lr){2-5}
& \textit{Domain average} & 0.410 & \textbf{0.513} & +0.103 \\
\midrule
\multirow{4}{*}{Proteins}
& Text-conditioned functional protein design & \textbf{0.586} & 0.522 & -0.064 \\
& Binder design & 0.297 & \textbf{0.645} & +0.348 \\
& Protein function prediction & 0.000 & \textbf{0.055} & +0.055 \\
\cmidrule(lr){2-5}
& \textit{Domain average} & 0.294 & \textbf{0.407} & +0.113 \\
\midrule
\multirow{3}{*}{Genomic sequences}
& DNA regulatory/splice span localization & 0.208 & \textbf{0.516} & +0.308 \\
& RNA family/anticodon span localization & 0.253 & \textbf{0.396} & +0.143 \\
\cmidrule(lr){2-5}
& \textit{Domain average} & 0.226 & \textbf{0.468} & +0.242 \\
\midrule
\multirow{4}{*}{Cells/pathways}
& Cell type recognition & 0.550 & \textbf{0.580} & +0.030 \\
& Hallmark program recognition & 0.700 & \textbf{0.750} & +0.050 \\
& Perturbation response prediction & \textbf{0.624} & 0.498 & -0.126 \\
\cmidrule(lr){2-5}
& \textit{Domain average} & \textbf{0.625} & 0.609 & -0.016 \\
\midrule
\rowcolor{aliceblue}
\textbf{Biological overall}
&
& 0.385
& \textbf{0.499}
& +0.114 \\
\midrule
\multirow{6}{*}{General language}
& MMLU (5-shot) & \textbf{61.8} & 59.7 & -2.1 \\
& ARC-c (10-shot) & \textbf{58.0} & 57.0 & -1.0 \\
& HellaSwag (10-shot) & \textbf{77.6} & 76.5 & -1.1 \\
& Winogrande (0-shot) & \textbf{70.6} & 70.5 & -0.1 \\
& PIQA (10-shot) & \textbf{81.7} & 81.1 & -0.6 \\
\cmidrule(lr){2-5}
& \textit{Average} & \textbf{69.9} & 69.0 & -0.9 \\
\bottomrule
\end{tabularx}
\end{table*}

To isolate the contribution of the \system{} corpus from the effect of annealing on general text alone, we introduce \textbf{Text-annealing only}, the model obtained by training the base checkpoint on just the general scientific and web text described above, with no \system{} data. Since text-annealing only and Ours share the same base checkpoint, the same annealing process, and the same general text, their only difference is the presence of the \system{} corpus, so any gap between them isolates the contribution of our biological data on downstream tasks (\cref{tab:textanneal-vs-ours}).

General text alone already lifts overall performance from 0.223 to 0.385. Much of this comes from scientific sources: the annealing mixture includes about 54B tokens of PubMed, bioRxiv, and medRxiv text, comparable in scale to \system{} itself (\cref{tab:scientific-text}), which already describes molecules, proteins, and pathways in prose. Annealing on it, therefore, transfers real biological signal, leaving the text-only model far above the base and even competitive where free text states facts explicitly, such as cross-domain reasoning (0.493).
Training on \system{} raises overall performance further to 0.499, with the clearest gains where free text falls short, on structured and tool-derived tasks: genomic localization improves by 0.242 and binder design by 0.348, with forward synthesis and molecule generation rising comparably. Cross-domain reasoning also improves, reaching 0.507 over the text-only model's 0.493. On text-conditioned functional protein design, the \system{} model scores slightly below the text-only model (0.522 vs 0.586), but the text-only model produces degenerate sequences (nondegeneracy 0.160 for binder generation, 0.540 for functional design), whereas \system{} generates diverse ones (1.000 and 0.930). Adding \system{} also lowers perturbation-response prediction, which we attribute to the thin cell-domain coverage of the current corpus; enriching this is left to future work.

\paragraph{Preservation of general-language ability.}
The gains above come from shifting the training mixture heavily toward biological data, which raises the concern that general-language ability may degrade in return. We check this by comparing both models on five standard language benchmarks~\citep{mmlu,arc,hellaswag,winogrande,piqa} in \cref{tab:textanneal-vs-ours}. Despite the biological corpus dominating the mixture, the model trained with \system{} stays within 0.9 points of the text-annealing only model on average (69.0 vs 69.9), and the largest single-benchmark drop is only 2.1 points, on MMLU. The broad biological capability that \system{} adds therefore comes with little forgetting, leaving general-language ability nearly intact.

\subsection{Cross-domain understanding}
\label{sec:cross_domain}


\begin{table}[t]
\centering
\footnotesize
\caption{Cross-domain two-hop reasoning: \textbf{Base} vs.\ \textbf{Ours}. \textbf{Base} is the base model checkpoint before training; \textbf{Ours} adds the \system{} corpus on top of general scientific and web text. Each subtask is a four-way multiple-choice question linking an entity to a knowledge-graph relation. Bold indicates the best model in each row.}
\label{tab:bmfm-cross-domain}

\setlength{\tabcolsep}{5pt}
\renewcommand{\arraystretch}{0.95}

\begin{tabularx}{0.92\linewidth}{
>{\raggedright\arraybackslash}X
r
r
r
}
\toprule
\textbf{Subtask} & \textbf{Base} & \textbf{Ours} & $\Delta$ \\
\midrule
Protein function $\rightarrow$ pathway & 0.420 & \textbf{0.680} & +0.260 \\
TF function $\rightarrow$ regulated target gene & 0.260 & \textbf{0.480} & +0.220 \\
Small molecule $\rightarrow$ binding target, pathway & 0.260 & \textbf{0.360} & +0.100 \\
\cmidrule(lr){1-4}
\textit{Average} & 0.313 & \textbf{0.507} & +0.194 \\
\bottomrule
\end{tabularx}
\end{table}

The suite in Section~\ref{sec:main_results} scores each biological domain in isolation. To test
whether ours yields genuinely cross-domain reasoning, we evaluate three two-hop relational tasks, each posed as four-way multiple choice. Every question begins from one entity's function or identity and requires a second hop into a knowledge-graph relationship: a protein's function to a pathway it participates in (SwissProt~$\rightarrow$~Reactome/Hetionet), a transcription factor's function to a gene it directly regulates (SwissProt~$\rightarrow$~OmniPath), and a small molecule to the (target protein, pathway) pair it acts through (DRKG~$\rightarrow$~Reactome/Hetionet). Entity names and gene symbols are withheld, so the first hop cannot be solved by surface lookup. 

We report the zero-shot accuracy in \cref{tab:bmfm-cross-domain}. Across all three relation types, the model trained on our corpus consistently outperforms the base model, raising the average accuracy from 0.313 to 0.507, with the largest improvements on protein function-to-pathway. This suggests that \system{} strengthens cross-domain knowledge integration, not just single-domain task performance.

\section{Related Work}
\subsection{Biological foundation language models}

A growing body of work places biological entities under a shared language interface, so one model can read and generate across molecules, proteins, and nucleic acids~\citep{biot5,biot5+,txllm,txgemma}. The most ambitious treat many entity types as one ``language'': NatureLM~\citep{naturelm} models small molecules, proteins, DNA, RNA, and materials as sequences in a single model and SciReasoner~\citep{scireasoner} aligns language with heterogeneous scientific representations and adds reasoning-oriented post-training. LOGOS~\citep{logos} extends this idea by representing small molecules, proteins, and material entities and their 3D spatial interactions through a shared text-based scientific grammar. These establish the language-interface paradigm \system{} builds on, but each centers on a model rather than a corpus, and stays at the molecular sequence levels.

\subsection{Modality-specific foundation models}

Outside the language-model paradigm, modality-specific models pair a pretrained model with a large domain corpus: ESM2~\citep{esm2} and ProGen2~\citep{progen2} for protein sequences, the Nucleotide Transformer~\citep{nucleotide_transformer} and DNABERT-2~\citep{dnabert2} for genomes, Geneformer~\citep{geneformer} and scGPT~\citep{scgpt} for single cells, and Evo~\citep{evo,evo2} for cross-modal genomic modeling. Each is powerful within its modality but does not span levels or expose entities through language.
\subsection{Biological corpora and benchmarks}

A parallel line releases the data and evaluations such models need. Instruction corpora such as Mol-Instructions~\citep{Mol-instructions} cover molecule, protein, and biomolecular-text tasks, while benchmark suites standardize evaluation within a domain: TAPE and PEER~\citep{tape,peer} and ProteinGym~\citep{proteingym} for proteins, GUE~\citep{dnabert2} and BEND~\citep{bend} for genomics, and the Therapeutics Data Commons~\citep{tdc} for drug discovery. These remain largely single-domain and tied to one or a few modalities. \system{} instead pairs a corpus with a matched evaluation suite across molecular, protein, genomic, cellular, and cross-domain settings, and goes beyond aggregation by enriching entries with tool-computed properties verbalized into language and by constructing new instruction data for practical tasks that existing resources underserve.
\section{Conclusion}
We introduced \system{}, an open pre-training scale corpus  for biology on training BioLM, spanning small molecules, proteins, genomic sequences, cells, and pathways. We convert diverse biological resources into a shared language interface and polish them into structured textual records. We further enrich these records with biological features computed by computational tools, and build instruction tasks for capabilities that conventional corpora underrepresent, such as protein binding, DNA/RNA feature localization, and cross-domain reasoning. In addition, we present \evalsuite{} to assess broad biological capabilities across domain-specific and cross-domain reasoning tasks. Holding the Gravity-16B-A3B architecture fixed, training a base model on \system{} substantially improves overall biological task performance across all domains. These results highlight that carefully constructed data can be the main driver of practical biological capability in BioLM.

\section*{Acknowledgments}

This work was partially supported by the Ministry of Science and ICT (MSIT), Republic of Korea, through the National IT Industry Promotion Agency (NIPA), as part of the Domain-Specific Foundation Model Project (Grant No. PJT-26-100004). We thank Mingu Kang, Cedric Caruzzo, Wonseok Lee, Steve Immanuel, Donggeun Yoo, Gihyeon Lee, Jin Woo Oh, Aisha Urooj, Laurent Dillard, and Aaron Valero from Lunit; Jeongwook Lee and Sunggi An from Seoul National University; Saebom Leem and Hyunkyu Jung from KAIST; and Beommo Kim from Aigen Sciences Inc. for their support and contributions throughout the project.

\bibliography{bib}

@STRING{NeurIPS = "Advances in Neural Information Processing Systems"}

@STRING{ICML = "International Conference on Machine Learning"}

@STRING{ICLR = "International Conference on Learning Representations"}

@STRING{ACL = "Annual Conference of the Association for Computational Linguistics"}

@STRING{EMNLP = "Conference on Empirical Methods in Natural Language Processing"}

@STRING{AAAI = "AAAI Conference on Artificial Intelligence"}

@article{vibeproteinbench,
  title={VibeProteinBench: An Evaluation Benchmark for Language-interfaced Vibe Protein Design},
  author={Seo, Hyunjin and Ahn, Hongjoon and Park, Jimin and Han, Sungjun and Lee, Gyubok and Yang, Soojung and Brown, Joseph S and Chen, Leo and Nesr, Gina El and Eweje, Feyisayo and others},
  journal={arXiv preprint arXiv:2605.10978},
  year={2026}
}

@article{naturelm,
  title={Nature language model: deciphering the language of nature for scientific discovery},
  author={Xia, Yingce and Jin, Peiran and Xie, Shufang and He, Liang and Cao, Chuan and Luo, Renqian and Liu, Guoqing and Wang, Yue and Liu, Zequn and Chen, Yuan-Jyue and others},
  journal={arXiv preprint arXiv:2502.07527},
  year={2025}
}

@article{scireasoner,
  title={SciReasoner: Laying the Scientific Reasoning Ground Across Disciplines},
  author={Wang, Yizhou and Tang, Chen and Deng, Han and Xiao, Jiabei and Liu, Jiaqi and Wu, Jianyu and Yao, Jun and Li, Pengze and Su, Encheng and Wang, Lintao and others},
  journal={arXiv preprint arXiv:2509.21320},
  year={2025}
}

@article{txgemma,
  title={Txgemma: Efficient and agentic llms for therapeutics},
  author={Wang, Eric and Schmidgall, Samuel and Jaeger, Paul F and Zhang, Fan and Pilgrim, Rory and Matias, Yossi and Barral, Joelle and Fleet, David and Azizi, Shekoofeh},
  journal={arXiv preprint arXiv:2504.06196},
  year={2025}
}

@article{Mol-instructions,
  title={Mol-instructions: A large-scale biomolecular instruction dataset for large language models},
  author={Fang, Yin and Liang, Xiaozhuan and Zhang, Ningyu and Liu, Kangwei and Huang, Rui and Chen, Zhuo and Fan, Xiaohui and Chen, Huajun},
  journal={arXiv preprint arXiv:2306.08018},
  year={2023}
}

@inproceedings{
mollangbench,
title={MolLangBench: A Comprehensive Benchmark for Language-Prompted Molecular Structure Recognition, Editing, and Generation},
author={Feiyang Cai and Jiahui Bai and Tao Tang and Guijuan He and Joshua Luo and Tianyu Zhu and Srikanth Pilla and Gang Li and Ling Liu and Feng Luo},
booktitle=ICLR,
year={2026},
}

@inproceedings{biot5,
  title={Biot5: Enriching cross-modal integration in biology with chemical knowledge and natural language associations},
  author={Pei, Qizhi and Zhang, Wei and Zhu, Jinhua and Wu, Kehan and Gao, Kaiyuan and Wu, Lijun and Xia, Yingce and Yan, Rui},
  booktitle=EMNLP,
  year={2023}
}

@inproceedings{biot5+,
  title={Biot5+: Towards generalized biological understanding with iupac integration and multi-task tuning},
  author={Pei, Qizhi and Wu, Lijun and Gao, Kaiyuan and Liang, Xiaozhuan and Fang, Yin and Zhu, Jinhua and Xie, Shufang and Qin, Tao and Yan, Rui},
  booktitle={Findings of the Association for Computational Linguistics},
  year={2024}
}

@article{txllm,
  title={Tx-llm: A large language model for therapeutics},
  author={Chaves, Juan Manuel Zambrano and Wang, Eric and Tu, Tao and Vaishnav, Eeshit Dhaval and Lee, Byron and Mahdavi, S Sara and Semturs, Christopher and Fleet, David and Natarajan, Vivek and Azizi, Shekoofeh},
  journal={arXiv preprint arXiv:2406.06316},
  year={2024}
}

@article{esm2,
  title={Evolutionary-scale prediction of atomic-level protein structure with a language model},
  author={Lin, Zeming and Akin, Halil and Rao, Roshan and Hie, Brian and Zhu, Zhongkai and Lu, Wenting and Smetanin, Nikita and Verkuil, Robert and Kabeli, Ori and Shmueli, Yaniv and others},
  journal={Science},
  volume={379},
  number={6637},
  pages={1123--1130},
  year={2023}
}

@article{nucleotide_transformer,
  title={Nucleotide transformer: building and evaluating robust foundation models for human genomics},
  author={Dalla-Torre, Hugo and Gonzalez, Liam and Mendoza-Revilla, Javier and Lopez Carranza, Nicolas and Grzywaczewski, Adam Henryk and Oteri, Francesco and Dallago, Christian and Trop, Evan and De Almeida, Bernardo P and Sirelkhatim, Hassan and others},
  journal={Nature Methods},
  volume={22},
  number={2},
  pages={287--297},
  year={2025}
}

@inproceedings{dnabert2,
  title={DNABERT-2: Efficient foundation model and benchmark for multi-species genomes},
  author={Zhou, Zhihan and Ji, Yanrong and Li, Weijian and Dutta, Pratik and Davuluri, Ramana and Liu, Han},
  booktitle=ICLR,
  year={2024}
}

@article{geneformer,
  title={Transfer learning enables predictions in network biology},
  author={Theodoris, Christina V and Xiao, Ling and Chopra, Anant and Chaffin, Mark D and Al Sayed, Zeina R and Hill, Matthew C and Mantineo, Helene and Brydon, Elizabeth M and Zeng, Zexian and Liu, X Shirley and others},
  journal={Nature},
  volume={618},
  number={7965},
  pages={616--624},
  year={2023}
}

@article{scgpt,
  title={scGPT: toward building a foundation model for single-cell multi-omics using generative AI},
  author={Cui, Haotian and Wang, Chloe and Maan, Hassaan and Pang, Kuan and Luo, Fengning and Duan, Nan and Wang, Bo},
  journal={Nature methods},
  volume={21},
  number={8},
  pages={1470--1480},
  year={2024}
}

@article{evo,
  title={Sequence modeling and design from molecular to genome scale with Evo},
  author={Nguyen, Eric and Poli, Michael and Durrant, Matthew G and Kang, Brian and Katrekar, Dhruva and Li, David B and Bartie, Liam J and Thomas, Armin W and King, Samuel H and Brixi, Garyk and others},
  journal={Science},
  volume={386},
  number={6723},
  pages={eado9336},
  year={2024}
}

@article{evo2,
  title={Genome modelling and design across all domains of life with Evo 2},
  author={Brixi, Garyk and Durrant, Matthew G and Ku, Jerome and Naghipourfar, Mohsen and Poli, Michael and Sun, Gwanggyu and Brockman, Greg and Chang, Daniel and Fanton, Alison and Gonzalez, Gabriel A and others},
  journal={Nature},
  volume={652},
  number={8112},
  pages={1349--1361},
  year={2026}
}

@inproceedings{tape,
  title={Evaluating protein transfer learning with TAPE},
  author={Rao, Roshan and Bhattacharya, Nicholas and Thomas, Neil and Duan, Yan and Chen, Peter and Canny, John and Abbeel, Pieter and Song, Yun},
  booktitle=NEURIPS,
  year={2019}
}

@inproceedings{peer,
  title={Peer: a comprehensive and multi-task benchmark for protein sequence understanding},
  author={Xu, Minghao and Zhang, Zuobai and Lu, Jiarui and Zhu, Zhaocheng and Zhang, Yangtian and Chang, Ma and Liu, Runcheng and Tang, Jian},
  booktitle=NEURIPS,
  year={2022}
}

@inproceedings{proteingym,
  title={Proteingym: Large-scale benchmarks for protein fitness prediction and design},
  author={Notin, Pascal and Kollasch, Aaron and Ritter, Daniel and Van Niekerk, Lood and Paul, Steffanie and Spinner, Han and Rollins, Nathan and Shaw, Ada and Orenbuch, Rose and Weitzman, Ruben and others},
  booktitle=NEURIPS,
  year={2023}
}

@article{bend,
  title={Bend: Benchmarking dna language models on biologically meaningful tasks},
  author={Marin, Frederikke Isa and Teufel, Felix and Horlacher, Marc and Madsen, Dennis and Pultz, Dennis and Winther, Ole and Boomsma, Wouter},
  journal={arXiv preprint arXiv:2311.12570},
  year={2023}
}

@article{tdc,
  title={Therapeutics data commons: Machine learning datasets and tasks for drug discovery and development},
  author={Huang, Kexin and Fu, Tianfan and Gao, Wenhao and Zhao, Yue and Roohani, Yusuf and Leskovec, Jure and Coley, Connor W and Xiao, Cao and Sun, Jimeng and Zitnik, Marinka},
  journal={arXiv preprint arXiv:2102.09548},
  year={2021}
}

@article{pubchem,
  title = {PubChem 2025 update},
  author = {Kim, Sunghwan and Chen, Jie and Cheng, Tiejun and Gindulyte, Asta and He, Jia and He, Siqian and Li, Qingliang and Shoemaker, Benjamin A. and Thiessen, Paul A. and Yu, Bo and Zaslavsky, Leonid and Zhang, Jian and Bolton, Evan E.},
  journal = {Nucleic Acids Research},
  volume = {53},
  number = {D1},
  pages = {D1516--D1525},
  year = {2025}
}

@article{bindingdb,
  title = {BindingDB in 2024: a FAIR knowledgebase of protein-small molecule binding data},
  author = {Liu, Tiqing and Hwang, Linda and Burley, Stephen K. and Nitsche, Carmen I. and Southan, Christopher and Walters, W. Patrick and Gilson, Michael K.},
  journal = {Nucleic Acids Research},
  volume = {53},
  number = {D1},
  pages = {D1633--D1644},
  year = {2025}
}

@article{chembl,
  title = {The ChEMBL Database in 2023: a drug discovery platform spanning multiple bioactivity data types and time periods},
  author = {Zdrazil, Barbara and Felix, Eloy and Hunter, Fiona and Manners, Emma J. and Blackshaw, James and Corbett, Sybilla and de Veij, Marleen and Ioannidis, Harris and Mendez Lopez, David and Mosquera, Juan F. and Magarinos, Maria Paula and Bosc, Nicolas and Arcila, Ricardo and Kiziloren, Tevfik and Gaulton, Anna and Bento, A. Patricia and Leach, Andrew R.},
  journal = {Nucleic Acids Research},
  volume = {52},
  number = {D1},
  pages = {D1180--D1192},
  year = {2024}
}

@article{uniprot,
  title = {UniProt: the Universal Protein Knowledgebase in 2025},
  author = {{The UniProt Consortium}},
  journal = {Nucleic Acids Research},
  volume = {53},
  number = {D1},
  pages = {D609--D617},
  year = {2025}
}

@article{pdb,
  title = {The Protein Data Bank},
  author = {Berman, Helen M. and Westbrook, John and Feng, Zukang and Gilliland, Gary and Bhat, T. N. and Weissig, Helge and Shindyalov, Ilya N. and Bourne, Philip E.},
  journal = {Nucleic Acids Research},
  volume = {28},
  number = {1},
  pages = {235--242},
  year = {2000}
}

@article{encode,
  title = {Expanded encyclopaedias of DNA elements in the human and mouse genomes},
  author = {Moore, Jill E. and Purcaro, Michael J. and Pratt, Henry E. and Epstein, Charles B. and Shoresh, Noam and Adrian, Jessika and Kawli, Trupti and Davis, Carrie A. and Dobin, Alexander and Kaul, Rajinder and others},
  journal = {Nature},
  volume = {583},
  number = {7818},
  pages = {699--710},
  year = {2020}
}

@article{gencode,
  title = {GENCODE 2021},
  author = {Frankish, Adam and Diekhans, Mark and Jungreis, Irwin and Lagarde, Julien and Loveland, Jane E. and Mudge, Jonathan M. and Sisu, Cristina and Wright, James C. and Armstrong, Joel and Barnes, If and others},
  journal = {Nucleic Acids Research},
  volume = {49},
  number = {D1},
  pages = {D916--D923},
  year = {2021}
}

@article{rnacentral,
  title = {RNAcentral 2021: secondary structure integration, improved sequence search and new member databases},
  author = {{The RNAcentral Consortium}},
  journal = {Nucleic Acids Research},
  volume = {49},
  number = {D1},
  pages = {D212--D220},
  year = {2021}
}

@article{rfam,
  title = {Rfam 14: expanded coverage of metagenomic, viral and microRNA families},
  author = {Kalvari, Ioanna and Nawrocki, Eric P. and Ontiveros-Palacios, Natalia and Argasinska, Joanna and Lamkiewicz, Kevin and Marz, Manja and Griffiths-Jones, Sam and Toffano-Nioche, Claire and Gautheret, Daniel and Weinberg, Zasha and others},
  journal = {Nucleic Acids Research},
  volume = {49},
  number = {D1},
  pages = {D192--D200},
  year = {2021}
}

@article{smolinstruct,
  title={Llasmol: Advancing large language models for chemistry with a large-scale, comprehensive, high-quality instruction tuning dataset},
  author={Yu, Botao and Baker, Frazier N and Chen, Ziqi and Ning, Xia and Sun, Huan},
  journal={arXiv preprint arXiv:2402.09391},
  year={2024}
}

@inproceedings{cell2sentence,
  title = {Cell2Sentence: Teaching Large Language Models the Language of Biology},
  author = {Levine, Daniel and Rizvi, Syed A. and L{\'e}vy, Sacha and Pallikkavaliyaveetil, Nazreen and Zhang, David and Chen, Xingyu and Ghadermarzi, Sina and Wu, Ruiming and Zheng, Zihe and Vrkic, Ivan and Zhong, Anna and Raskin, Daphne and Han, Insu and de Oliveira Fonseca, Antonio Henrique and Caro, Josue Ortega and Karbasi, Amin and Dhodapkar, Rahul Madhav and van Dijk, David},
  booktitle = ICML,
  year = {2024},
}

@article{tabulasapiens,
  title = {The Tabula Sapiens: A multiple-organ, single-cell transcriptomic atlas of humans},
  author = {{The Tabula Sapiens Consortium}},
  journal = {Science},
  volume = {376},
  number = {6594},
  pages = {eabl4896},
  year = {2022}
}

@inproceedings{perturbench,
  title={Perturbench: Benchmarking machine learning models for cellular perturbation analysis},
  author={Wu, Yan and Wershof, Esther and Schmon, Sebastian and Nassar, Marcel and Osi{\'n}ski, B{\l}a{\.z}ej and Eksi, Ridvan and Yan, Zichao and Stark, Rory and Zhang, Kun and Graepel, Thore},
  booktitle = NEURIPS,
  year={2026}
}

@article{hubmap,
  title = {Human BioMolecular Atlas Program (HuBMAP): 3D Human Reference Atlas construction and usage},
  author = {B{\"o}rner, Katy and Blood, Philip D. and Silverstein, Jonathan C. and others},
  journal = {Nature Methods},
  volume = {22},
  number = {4},
  pages = {845--860},
  year = {2025}
}

@article{jumpcellpainting,
  title = {Three million images and morphological profiles of cells treated with matched chemical and genetic perturbations},
  author = {Chandrasekaran, Srinivas Niranj and Cimini, Beth A. and Goodale, Amy and others},
  journal = {Nature Methods},
  volume = {21},
  pages = {1114--1121},
  year = {2024}
}

@article{go,
  title = {The Gene Ontology knowledgebase in 2023},
  author = {{The Gene Ontology Consortium}},
  journal = {Genetics},
  volume = {224},
  number = {1},
  pages = {iyad031},
  year = {2023}
}

@article{reactome,
  title = {The Reactome Pathway Knowledgebase 2024},
  author = {Milacic, Marija and Beavers, Deidre and Conley, Patrick and Gong, Chuqiao and Gillespie, Marc and Griss, Johannes and Haw, Robin and Jassal, Bijay and Matthews, Lisa and May, Bruce and others},
  journal = {Nucleic Acids Research},
  volume = {52},
  number = {D1},
  pages = {D672--D678},
  year = {2024}
}

@article{hetionet,
  title = {Systematic integration of biomedical knowledge prioritizes drugs for repurposing},
  author = {Himmelstein, Daniel S. and Lizee, Antoine and Hessler, Christine and Brueggeman, Leo and Chen, Sabrina L. and Hadley, Dexter and Green, Ari and Khankhanian, Pouya and Baranzini, Sergio E.},
  journal = {eLife},
  volume = {6},
  pages = {e26726},
  year = {2017}
}

@misc{drkg,
  title = {Drug Repurposing Knowledge Graph (DRKG)},
  author = {{Drug Repurposing Knowledge Graph Team}},
  year = {2020},
  howpublished = {\url{https://github.com/gnn4dr/DRKG}},
  note = {Accessed: 2026-06-12}
}

@article{string,
  title = {The STRING database in 2023: protein-protein association networks and functional enrichment analyses for any sequenced genome of interest},
  author = {Szklarczyk, Damian and Kirsch, Rebecca and Koutrouli, Mikaela and Nastou, Katerina and Mehryary, Farrokh and Hachilif, Radja and Gable, Annika L. and Fang, Tao and Doncheva, Nadezhda T. and Pyysalo, Sampo and Bork, Peer and Jensen, Lars J. and von Mering, Christian},
  journal = {Nucleic Acids Research},
  volume = {51},
  number = {D1},
  pages = {D638--D646},
  year = {2023},
}

@article{dorothea,
  title = {Benchmark and integration of resources for the estimation of human transcription factor activities},
  author = {Garcia-Alonso, Luz and Holland, Christian H. and Ibrahim, Mahmoud M. and Turei, Denes and Saez-Rodriguez, Julio},
  journal = {Genome Research},
  volume = {29},
  number = {8},
  pages = {1363--1375},
  year = {2019}
}

@article{humanproteinatlas,
  title = {Proteomics. Tissue-based map of the human proteome},
  author = {Uhl{\'e}n, Mathias and Fagerberg, Linn and Hallstr{\"o}m, Bj{\"o}rn M. and Lindskog, Cecilia and Oksvold, Per and Mardinoglu, Adil and Sivertsson, {\AA}sa and Kampf, Caroline and Sj{\"o}stedt, Evelina and Asplund, Anna and others},
  journal = {Science},
  volume = {347},
  number = {6220},
  pages = {1260419},
  year = {2015}
}

@inproceedings{proteinlmbench,
  title = {A Fine-tuning Dataset and Benchmark for Large Language Models for Protein Understanding},
  author = {Shen, Yiqing and Chen, Zan and Mamalakis, Michail and He, Luhan and Xia, Haiyang and Li, Tianbin and Su, Yanzhou and He, Junjun and Wang, Yu Guang},
  booktitle = {2024 IEEE International Conference on Bioinformatics and Biomedicine (BIBM)},
  year = {2024},
}

@article{bioreasonpro,
  title={BioReason-Pro: Advancing Protein Function Prediction with Multimodal Biological Reasoning},
  author={Fallahpour, Adibvafa and Seyed-Ahmadi, Arman and Idehpour, Parsa and Ibrahim, Omar and Gupta, Purav and Naimer, Jack and Zhu, Kevin and Shah, Arnav and Ma, Shihao and Adduri, Abhinav and others},
  journal={bioRxiv},
  pages={2026--03},
  year={2026},
}

@article{dssp,
  title={Dictionary of protein secondary structure: pattern recognition of hydrogen-bonded and geometrical features},
  author={Kabsch, Wolfgang and Sander, Chris},
  journal={Biopolymers},
  volume={22},
  number={12},
  pages={2577--2637},
  year={1983},
}

@article{afdb,
  title = {AlphaFold Protein Structure Database 2025: a redesigned interface and updated structural coverage},
  author = {Bertoni, Damian and Tsenkov, Maxim and Magana, Paulyna and Nair, Sreenath and Pidruchna, Ivanna and Querino Lima Afonso, Marcelo and Midlik, Adam and Paramval, Urmila and Lawal, Dare and Tanweer, Ahsan and Last, Meera and Patel, Risha and Laydon, Agata and Lasecki, Dariusz and Dietrich, Nick and Tomlinson, Hamish and {\v Z}{\'i}dek, Augustin and Green, Tim and Kovalevskiy, Oleg and Lau, Andy and Kandathil, Shaun and Bordin, Nicola and Sillitoe, Ian and Mirdita, Milot and Jones, David and Orengo, Christine and Steinegger, Martin and Fleming, Jennifer R. and Velankar, Sameer},
  journal = {Nucleic Acids Research},
  volume = {54},
  number = {D1},
  pages = {D358--D362},
  year = {2026},
}

@article{drugbank,
  title = {DrugBank 6.0: the DrugBank Knowledgebase for 2024},
  author = {Knox, Craig and Wilson, Mike and Klinger, Christen M. and others},
  journal = {Nucleic Acids Research},
  volume = {52},
  number = {D1},
  pages = {D1265--D1275},
  year = {2024},
}

@article{lincs_l1000,
  title = {A Next Generation Connectivity Map: L1000 Platform and the First 1,000,000 Profiles},
  author = {Subramanian, Aravind and Narayan, Rajiv and Corsello, Steven M. and others},
  journal = {Cell},
  volume = {171},
  number = {6},
  pages = {1437--1452.e17},
  year = {2017},
}

@article{cellxgene,
  title = {{CZ CELLxGENE} Discover: a single-cell data platform for scalable exploration, analysis and modeling of aggregated data},
  author = {{CZI Cell Science Program} and Abdulla, Shibla and Aevermann, Brian and others},
  journal = {Nucleic Acids Research},
  volume = {53},
  number = {D1},
  pages = {D886--D900},
  year = {2025},
}

@article{plinder,
  title={PLINDER: The protein-ligand interactions dataset and evaluation resource},
  author={Durairaj, Janani and Adeshina, Yusuf and Cao, Zhonglin and Zhang, Xuejin and Oleinikovas, Vladas and Duignan, Thomas and McClure, Zachary and Robin, Xavier and Studer, Gabriel and Kovtun, Daniel and others},
  journal={BioRxiv},
  year={2024},
}

@article{pinder,
  title={PINDER: The protein interaction dataset and evaluation resource},
  author={Kovtun, Daniel and Akdel, Mehmet and Goncearenco, Alexander and Zhou, Guoqing and Holt, Graham and Baugher, David and Lin, Dejun and Adeshina, Yusuf and Castiglione, Thomas and Wang, Xiaoyun and others},
  journal={bioRxiv},
  year={2024},
}

@inproceedings{ppiref,
  title = {Learning to Design Protein-Protein Interactions with Enhanced Generalization},
  author = {Bushuiev, Anton and Bushuiev, Roman and Kouba, Petr and Filkin, Anatolii and Gabrielova, Marketa and Gabriel, Michal and Sedlar, Jiri and Pluskal, Tomas and Damborsky, Jiri and Mazurenko, Stanislav and Sivic, Josef},
  booktitle = ICLR,
  year = {2024},
}

@article{dipsplus,
  title = {{DIPS-Plus}: The enhanced database of interacting protein structures for interface prediction},
  author = {Morehead, Alex and Chen, Chen and Sedova, Ada and Cheng, Jianlin},
  journal = {Scientific Data},
  volume = {10},
  number = {1},
  pages = {509},
  year = {2023},
}

@article{ppikb,
  title = {{PPIKB}: A Comprehensive Knowledge Base and Analysis Platform for Protein-Peptide Interactions Based on Literature and Patents},
  author = {Zhu, Ning and Ming, Yanyu and Zhang, Chengyun and Sen, Cao and Li, Chongyang and Guo, Jingjing and Duan, Hongliang},
  journal = {bioRxiv},
  year = {2025},
}

@article{propedia,
  title = {Propedia v2.3: A novel representation approach for the peptide-protein interaction database using graph-based structural signatures},
  author = {Martins, Pedro and Mariano, Diego and Carvalho, Frederico Chaves and others},
  journal = {Frontiers in Bioinformatics},
  volume = {3},
  year = {2023},
}

@article{replogle,
  title = {Mapping information-rich genotype-phenotype landscapes with genome-scale Perturb-seq},
  author = {Replogle, Joseph M. and Saunders, Reuben A. and Pogson, Angela N. and others},
  journal = {Cell},
  volume = {185},
  number = {14},
  pages = {2559--2575.e28},
  year = {2022},
}

@article{norman,
  title = {Exploring genetic interaction manifolds constructed from rich single-cell phenotypes},
  author = {Norman, Thomas M. and Horlbeck, Max A. and Replogle, Joseph M. and others},
  journal = {Science},
  volume = {365},
  number = {6455},
  pages = {786--793},
  year = {2019},
}

@article{jaspar,
  title = {{JASPAR} 2026: expansion of transcription factor binding profiles and integration of deep learning models},
  author = {Ovek Baydar, Damla and Rauluseviciute, Ieva and Aronsen, Dina R. and others},
  journal = {Nucleic Acids Research},
  volume = {54},
  number = {D1},
  pages = {D184--D193},
  year = {2026},
}

@article{mirbase,
  title = {{miRBase}: from microRNA sequences to function},
  author = {Kozomara, Ana and Birgaoanu, Maria and Griffiths-Jones, Sam},
  journal = {Nucleic Acids Research},
  volume = {47},
  number = {D1},
  pages = {D155--D162},
  year = {2019},
}

@article{gtrnadb,
  title = {{GtRNAdb} 2.0: an expanded database of transfer {RNA} genes identified in complete and draft genomes},
  author = {Chan, Patricia P. and Lowe, Todd M.},
  journal = {Nucleic Acids Research},
  volume = {44},
  number = {D1},
  pages = {D184--D189},
  year = {2016},
}

@inproceedings{flan_collection,
  title={The flan collection: Designing data and methods for effective instruction tuning},
  author={Longpre, Shayne and Hou, Le and Vu, Tu and Webson, Albert and Chung, Hyung Won and Tay, Yi and Zhou, Denny and Le, Quoc V and Zoph, Barret and Wei, Jason and others},
  booktitle=ICML,
  year={2023},
}

@misc{rdkit,
  title = {{rdkit/rdkit}: 2025\_09\_5 (Q3 2025) Release},
  author = {Landrum, Greg and Tosco, Paolo and Kelley, Brian and Rodriguez, Ricardo and Cosgrove, David and Vianello, Riccardo and Gedeck, Peter and others},
  year = {2026},
  publisher = {Zenodo},
  doi = {10.5281/zenodo.18428170},
  url = {https://doi.org/10.5281/zenodo.18428170},
  note = {Version Release\_2025\_09\_5}
}

@article{primer,
  title={Language models for biological research: a primer},
  author={Simon, Elana and Swanson, Kyle and Zou, James},
  journal={Nature Methods},
  volume={21},
  number={8},
  pages={1422--1429},
  year={2024},
  publisher={Nature Publishing Group US New York}
}

@article{omnipath,
  title={OmniPath: guidelines and gateway for literature-curated signaling pathway resources},
  author={T{\"u}rei, D{\'e}nes and Korcsm{\'a}ros, Tam{\'a}s and Saez-Rodriguez, Julio},
  journal={Nature methods},
  volume={13},
  number={12},
  pages={966--967},
  year={2016},
  publisher={Nature Publishing Group US New York}
}

@article{esmfold2,
  title={Language Modeling Materializes a World Model of Protein Biology},
  author={Candido, Salvatore and Hayes, Thomas and Derry, Alexander and Rao, Roshan and Lin, Zeming and Verkuil, Robert and Wu, Bryan Z and Lee, Jin Sub and Bruguera, Elise S and Keval, Jehan A and others},
  journal={bioRxiv},
  pages={2026--06},
  year={2026},
}

@misc{gravity-moe-2026,
    title={Gravity-16B-A3B-Base},
    author={{Trillion Labs}},
    year={2026},
    url={https://huggingface.co/trillionlabs/Gravity-16B-A3B-Base}
}

@article{sctype,
  title = {Fully-automated and ultra-fast cell-type identification using specific marker combinations from single-cell transcriptomic data},
  author = {Ianevski, Aleksandr and Giri, Anil K. and Aittokallio, Tero},
  journal = {Nature Communications},
  volume = {13},
  number = {1},
  pages = {1246},
  year = {2022},
}

@article{msigdb_hallmark,
  title = {The Molecular Signatures Database Hallmark Gene Set Collection},
  author = {Liberzon, Arthur and Birger, Chet and Thorvaldsd{\'o}ttir, Helga and Ghandi, Mahmoud and Mesirov, Jill P. and Tamayo, Pablo},
  journal = {Cell Systems},
  volume = {1},
  number = {6},
  pages = {417--425},
  year = {2015},
}

@article{enrichr,
  title = {Enrichr: interactive and collaborative HTML5 gene list enrichment analysis tool},
  author = {Chen, Edward Y. and Tan, Christopher M. and Kou, Yan and Duan, Qiaonan and Wang, Zichen and Meirelles, Gabriela Vaz and Clark, Neil R. and Ma'ayan, Avi},
  journal = {BMC Bioinformatics},
  volume = {14},
  pages = {128},
  year = {2013},
}

@article{mazein2024graph,
  title={Graph databases in systems biology: a systematic review},
  author={Mazein, Ilya and Rougny, Adrien and Mazein, Alexander and Henkel, Ron and G{\"u}tebier, Lea and Michaelis, Lea and Ostaszewski, Marek and Schneider, Reinhard and Satagopam, Venkata and Jensen, Lars Juhl and others},
  journal={Briefings in Bioinformatics},
  volume={25},
  number={6},
  pages={bbae561},
  year={2024},
}

@article{mohamed2021biological,
  title={Biological applications of knowledge graph embedding models},
  author={Mohamed, Sameh K and Nounu, Aayah and Nov{\'a}{\v{c}}ek, V{\'\i}t},
  journal={Briefings in bioinformatics},
  volume={22},
  number={2},
  pages={1679--1693},
  year={2021},
}

@article{jang2026towards,
  title={Towards Autonomous Mechanistic Reasoning in Virtual Cells},
  author={Jang, Yunhui and Zhu, Lu and Fawkes, Jake and Denton, Alisandra Kaye and Beaini, Dominique and Noutahi, Emmanuel},
  journal={arXiv preprint arXiv:2604.11661},
  year={2026}
}

@article{chatnt,
  title={Chatnt: A multimodal conversational agent for dna, rna and protein tasks},
  author={Richard, Guillaume and De Almeida, Bernardo P and Dalla-Torre, Hugo and Blum, Christopher and Hexemer, Lorenz and Pandey, Priyanka and Laurent, Stefan and Lopez, Marie and Laterre, Alexandre and Lang, Maren and others},
  journal={bioRxiv},
  pages={2024--04},
  year={2024},
}

@article{uspto50k,
  title = {What's What: The (Nearly) Definitive Guide to Reaction Role Assignment},
  author = {Schneider, Nadine and Stiefl, Nikolaus and Landrum, Gregory A.},
  journal = {Journal of Chemical Information and Modeling},
  volume = {56},
  number = {12},
  pages = {2336--2346},
  year = {2016},
}

@article{chebi20,
  title={A Quantitative Analysis of Knowledge-Learning Preferences in Large Language Models in Molecular Science},
  author={Liu, Pengfei and Tao, Jun and Ren, Zhixiang},
  journal={arXiv preprint arXiv:2402.04119},
  year={2024}
}

@misc{chemdata700k,
  title={ChemLLM: A Chemical Large Language Model},
  author={Zhang, Di and Liu, Wei and Tan, Qian and Chen, Jingdan and Yan, Hang and Yan, Yuliang and Li, Jiatong and Huang, Weiran and Yue, Xiangyu and Zhou, Dongzhan and Zhang, Shufei and Su, Mao and Zhong, Han-Sen and Li, Yuqiang and Ouyang, Wanli},
  year={2024},
  eprint={2402.06852},
  archivePrefix={arXiv},
  primaryClass={cs.AI}
}

@inproceedings{lpm24,
  title={{L}+{M}-24: Building a Dataset for {L}anguage+{M}olecules @ {ACL} 2024},
  author={Edwards, Carl and Wang, Qingyun and Zhao, Lawrence and Ji, Heng},
  booktitle={Proceedings of the 1st Workshop on Language + Molecules},
  pages={1--9},
  year={2024},
  address={Bangkok, Thailand},
  publisher={Association for Computational Linguistics},
  url={https://aclanthology.org/2024.langmol-1.1},
}

@inproceedings{generif,
  title={Gene Indexing: Characterization and Analysis of NLM's GeneRIFs},
  author={Mitchell, Joyce A. and Aronson, Alan R. and Mork, James G. and Folk, L. C. and Humphrey, Susanne M. and Ward, Janice M.},
  booktitle={AMIA Annual Symposium Proceedings},
  pages={460--464},
  year={2003}
}

@article{ncbi_gene,
  title={Entrez Gene: gene-centered information at NCBI},
  author={Maglott, Donna and Ostell, Jim and Pruitt, Kim D and Tatusova, Tatiana},
  journal={Nucleic acids research},
  volume={33},
  number={suppl\_1},
  pages={D54--D58},
  year={2005},
}

@article{cell_ontology,
  title={The Cell Ontology 2016: Enhanced Content, Modularization, and Ontology Interoperability},
  author={Diehl, Alexander D. and Meehan, Terrence F. and Bradford, Yvonne M. and Brush, Matthew H. and Dahdul, Wasila M. and Dougall, David S. and He, Yongqun and Osumi-Sutherland, David and Ruttenberg, Alan and Sarntivijai, Sirarat and Van Slyke, Ceri E. and Vasilevsky, Nicole A. and Haendel, Melissa A. and Blake, Judith A. and Mungall, Christopher J.},
  journal={Journal of Biomedical Semantics},
  volume={7},
  pages={44},
  year={2016},
}

@article{dgidb,
  title={Integration of the Drug--Gene Interaction Database ({DGIdb} 4.0) with Open Crowdsource Efforts},
  author={Freshour, Sharon L. and Kiwala, Susanna and Cotto, Kelsy C. and Coffman, Adam C. and McMichael, Joshua F. and Song, Jing and Griffith, Malachi and Griffith, Obi L. and Wagner, Alex H.},
  journal={Nucleic Acids Research},
  volume={49},
  number={D1},
  pages={D1144--D1151},
  year={2021},
}

@article{biopython,
  title={Biopython: freely available Python tools for computational molecular biology and bioinformatics},
  author={Cock, Peter JA and Antao, Tiago and Chang, Jeffrey T and Chapman, Brad A and Cox, Cymon J and Dalke, Andrew and Friedberg, Iddo and Hamelryck, Thomas and Kauff, Frank and Wilczynski, Bartek and de Hoon, Michiel JL},
  journal={Bioinformatics},
  volume={25},
  number={11},
  pages={1422--1423},
  year={2009},
}

@article{genbank,
  title   = {GenBank 2025 update},
  author  = {Sayers, Eric W. and Cavanaugh, Mark and Frisse, Linda and Pruitt, Kim D. and Schneider, Valerie A. and Underwood, Beverly A. and Yankie, Linda and Karsch-Mizrachi, Ilene},
  journal = {Nucleic Acids Research},
  volume  = {53},
  number  = {D1},
  pages   = {D56--D61},
  year    = {2025},
  doi     = {10.1093/nar/gkae1114}
}

@article{pyfaidx,
  title   = {Efficient ``pythonic'' access to FASTA files using pyfaidx},
  author  = {Shirley, Matthew D. and Ma, Zhaorong and Pedersen, Brent S. and Wheelan, Sarah J.},
  journal = {PeerJ PrePrints},
  volume  = {3},
  pages   = {e970v1},
  year    = {2015},
  doi     = {10.7287/peerj.preprints.970v1}
}

@article{anndata,
  title   = {anndata: Access and store annotated data matrices},
  author  = {Virshup, Isaac and Rybakov, Sergei and Theis, Fabian J. and Angerer, Philipp and Wolf, F. Alexander},
  journal = {Journal of Open Source Software},
  volume  = {9},
  number  = {101},
  pages   = {4371},
  year    = {2024},
  doi     = {10.21105/joss.04371}
}

@article{scipy,
  title   = {{SciPy} 1.0: fundamental algorithms for scientific computing in {Python}},
  author  = {Virtanen, Pauli and Gommers, Ralf and Oliphant, Travis E. and Haberland, Matt and Reddy, Tyler and Cournapeau, David and others},
  journal = {Nature Methods},
  volume  = {17},
  pages   = {261--272},
  year    = {2020},
  doi     = {10.1038/s41592-019-0686-2}
}

@article{numpy,
  title   = {Array programming with {NumPy}},
  author  = {Harris, Charles R. and Millman, K. Jarrod and van der Walt, St{\'e}fan J. and Gommers, Ralf and Virtanen, Pauli and Cournapeau, David and others},
  journal = {Nature},
  volume  = {585},
  pages   = {357--362},
  year    = {2020},
  doi     = {10.1038/s41586-020-2649-2}
}

@inproceedings{pandas,
  title     = {Data structures for statistical computing in {Python}},
  author    = {McKinney, Wes},
  booktitle = {Proceedings of the 9th Python in Science Conference},
  pages     = {56--61},
  year      = {2010},
  doi       = {10.25080/Majora-92bf1922-00a}
}

@article{sklearn,
  title   = {Scikit-learn: Machine learning in {Python}},
  author  = {Pedregosa, Fabian and Varoquaux, Ga{\"e}l and Gramfort, Alexandre and Michel, Vincent and Thirion, Bertrand and Grisel, Olivier and others},
  journal = {Journal of Machine Learning Research},
  volume  = {12},
  pages   = {2825--2830},
  year    = {2011}
}

@article{cellprofiler,
  title   = {CellProfiler: image analysis software for identifying and quantifying cell phenotypes},
  author  = {Carpenter, Anne E. and Jones, Thouis R. and Lamprecht, Michael R. and Clarke, Colin and Kang, In Han and Friman, Ola and others},
  journal = {Genome Biology},
  volume  = {7},
  pages   = {R100},
  year    = {2006},
  doi     = {10.1186/gb-2006-7-10-r100}
}

@inproceedings{mmlu,
  title={Measuring Massive Multitask Language Understanding},
  author={Hendrycks, Dan and Burns, Collin and Basart, Steven and Zou, Andy and Mazeika, Mantas and Song, Dawn and Steinhardt, Jacob},
  booktitle={International Conference on Learning Representations},
  year={2021}
}

@article{arc,
  title={Think you have solved question answering? try arc, the ai2 reasoning challenge},
  author={Clark, Peter and Cowhey, Isaac and Etzioni, Oren and Khot, Tushar and Sabharwal, Ashish and Schoenick, Carissa and Tafjord, Oyvind},
  journal={arXiv preprint arXiv:1803.05457},
  year={2018}
}

@inproceedings{hellaswag,
  title={Hellaswag: Can a machine really finish your sentence?},
  author={Zellers, Rowan and Holtzman, Ari and Bisk, Yonatan and Farhadi, Ali and Choi, Yejin},
  booktitle={Proceedings of the 57th annual meeting of the association for computational linguistics},
  pages={4791--4800},
  year={2019}
}

@article{winogrande,
  title={Winogrande: An adversarial winograd schema challenge at scale},
  author={Sakaguchi, Keisuke and Bras, Ronan Le and Bhagavatula, Chandra and Choi, Yejin},
  journal={Communications of the ACM},
  volume={64},
  number={9},
  pages={99--106},
  year={2021},
  publisher={ACM New York, NY, USA}
}

@inproceedings{piqa,
  title={Piqa: Reasoning about physical commonsense in natural language},
  author={Bisk, Yonatan and Zellers, Rowan and Gao, Jianfeng and Choi, Yejin and others},
  booktitle={Proceedings of the AAAI conference on artificial intelligence},
  volume={34},
  number={05},
  pages={7432--7439},
  year={2020}
}

@article{progen2,
  title={Progen2: exploring the boundaries of protein language models},
  author={Nijkamp, Erik and Ruffolo, Jeffrey A and Weinstein, Eli N and Naik, Nikhil and Madani, Ali},
  journal={Cell systems},
  volume={14},
  number={11},
  pages={968--978},
  year={2023},
  publisher={Elsevier}
}

@article{logos,
  title={Speaking the Language of Science: Toward a General-Purpose Generative Foundation Model for the Natural Sciences},
  author={Li, Mingyang and Liu, Yurou and Ye, Jieping and Su, Bing and Wen, Ji-Rong and Wang, Zheng},
  journal={arXiv preprint arXiv:2606.16905},
  year={2026}
}

\newpage
\appendix
\renewcommand{\sectionautorefname}{Appendix}
\renewcommand{\subsectionautorefname}{Appendix} 

\crefalias{section}{appendix}
\section*{Appendix}

\section{Dataset Collection Details}\label{A}

We collect existing biological resources that already encode useful molecular, protein, genomic, cells, and cross-domain knowledge, and normalize them into a shared corpus schema. 


\paragraph{Small molecules.} We draw on large public chemical and bioactivity resources. PubChem~\citep{pubchem} provides broad chemical identity and property coverage, DrugBank~\citep{drugbank} contributes drug-level annotations such as development stage and mechanism of action, and BindingDB~\citep{bindingdb} and ChEMBL~\citep{chembl} supply protein--ligand bioactivity and binding measurements. USPTO-50K~\citep{uspto50k} provides reaction records for forward-synthesis supervision.


\paragraph{Proteins.} We collect protein sequences, structures, and annotations from established repositories. UniProt/SwissProt~\citep{uniprot} provides curated sequences and functional annotations, while the AlphaFold database~\citep{afdb} and the Protein Data Bank (PDB)~\citep{pdb} provide predicted and experimental structures.


\paragraph{Genomic sequences.} We integrate established annotation resources. ENCODE~\citep{encode} provides regulatory annotations and assay-derived evidence over functional genomic elements, GENCODE~\citep{gencode} supplies reference gene and transcript annotations, and RNAcentral and Rfam~\citep{rnacentral,rfam} provide non-coding RNA sequences and RNA family annotations.


\paragraph{Cells and pathways.} We include resources that describe cellular states, expression, perturbation responses, and pathway-level biology. L1000~\citep{lincs_l1000} provides large-scale transcriptional perturbation profiles, and the Human Protein Atlas~\citep{humanproteinatlas} contributes single-cell-type expression data. GeneRIF~\citep{generif} and NCBI Gene~\citep{ncbi_gene} supply gene-level functional summaries, while Gene Ontology/GOA~\citep{go}, Reactome~\citep{reactome}, and the Cell Ontology~\citep{cell_ontology} provide functional, pathway, and cell-type annotations. We also draw on CELLxGENE~\citep{cellxgene} for aggregated single-cell data.

\paragraph{Cross-domain knowledge graphs.} To link entities across domains, we integrate biomedical knowledge graphs that encode typed relations among compounds, proteins, genes, pathways, and cell contexts: Hetionet~\citep{hetionet}, DRKG~\citep{drkg}, DGIdb~\citep{dgidb}, STRING~\citep{string}, OmniPath~\citep{omnipath}, and Reactome~\citep{reactome}.

\section{Tool-generated Dataset Details}\label{B}

\paragraph{Small molecules.} For small molecules, we convert PubChem records into computationally enriched molecule narratives using RDKit~\citep{rdkit}. We first parse each PubChem SMILES string with RDKit, canonicalize the molecular graph when parsing succeeds, and retain source-grounded identifiers and descriptions such as SMILES, IUPAC name, InChI, molecular formula, atom and bond counts, molecular weight, exact mass, charge state, hydrogen-bond donor/acceptor counts, rotatable bonds, and reported logP values. We then add deterministic RDKit-derived structural features that are rarely available in ordinary free-text descriptions, including ring counts and ring subtypes, aromatic/aliphatic/saturated ring composition, carbocyclic and heterocyclic ring counts, heteroatom count, amide bonds, spiro and bridgehead atoms, stereocenter counts, fraction of sp3 carbons, Murcko scaffold, Bertz complexity, Labute ASA, Hall--Kier alpha, kappa shape indices, valence and radical electron counts, functional-group matches, PAINS/Brenk alerts, and Morgan/MACCS fingerprint summaries. These computed properties are rendered into natural-language descriptions with explicit provenance, turning machine-computable molecular graph information into language-facing pretraining examples while preserving the original PubChem molecular identity.

\paragraph{Proteins.} We generate structure, binding, and function narratives from UniRef50/UniProt-linked sequences and structures, using AlphaFold database (AFDB)~\citep{afdb} and Protein Data Bank (PDB)~\citep{pdb} as structure sources. We extract secondary structure and backbone hydrogen-bond annotations using DSSP~\citep{dssp}, FreeSASA for solvent accessibility, and rule-based geometric analyses for Ramachandran regions, contact maps, long-range contacts, salt bridges, disulfide bonds, $\pi$-stacking, cation-$\pi$ interactions, radius of gyration, and relative contact order. The resulting structure narratives describe each protein's sequence, residue length, structure source, physicochemical properties, structural confidence, secondary-structure composition and residue spans, fold topology, compactness, interaction network, solvent exposure, residue composition, motifs, domains, GO annotations, active sites, and low-confidence or potentially disordered regions. We further construct binding narratives from protein--protein, protein--ligand, and protein--peptide complex resources, rendering interface residues and partner context as language-facing examples. Together, these sources expose proteins as physical, functional, and interaction-domain entities, rather than only amino-acid strings or free-text mentions.

\paragraph{Genomic sequences.} We generate narrative records from regulatory DNA intervals and RNA transcript annotations, rather than treating nucleotide sequences as plain strings. For DNA, we use GENCODE~\citep{gencode} genome annotations and ENCODE~\citep{encode} cCRE/peak resources to construct sequence-grounded regulatory narratives over genomic intervals. Each record combines the raw sequence with coordinate metadata, organism and assembly information, regulatory class or assay evidence when available, and deterministic sequence features such as length, GC/AT content, CpG count and observed-to-expected ratio, entropy, homopolymer runs, palindromic k-mers, ambiguous-base content, and simple composition flags. For RNA, we build transcript-domain narratives from GENCODE transcript sequences and related RNA resources~\citep{rnacentral,rfam}, adding computable features such as base composition, dinucleotide counts, GC/AU skew, AUG and stop-codon counts, ORF statistics, polyadenylation signal motifs, U-rich windows, and deterministic stem-loop candidates. 

\paragraph{Cells.} We convert cellular profile resources into language-facing records that capture expression, spatial context, perturbation response, and morphology. Single-cell records describe cell states from ranked or highly expressed marker genes and associated tissue context, while perturbation records connect CRISPRa/CRISPRi perturbations to pseudobulk differential-expression signatures. For spatial biology, we generate HuBMAP-derived spatial transcriptomics narratives~\citep{hubmap} that attach each spot to its tissue, assay, coordinate, detected-gene count, total transcript count, local neighborhood density, nearest-neighbor distance, and locally abundant transcripts. For morphology, we generate JUMP Cell Painting narratives~\citep{jumpcellpainting} from matched-control-normalized CellProfiler profiles, summarizing perturbation-domain phenotype strength, replicate consistency, morphology-family shifts across nuclei, cytoplasm, and whole cells, and the top contributing image-derived measurements. 

\paragraph{Cross-domain linking.} Cross-domain KG-linked records are constructed as chain-style training examples that co-locate evidence across biological domains. We use a knowledge-graph structure to decide which entities should appear together, then hydrate those entities with payloads from the original databases. Gene-anchored chains combine UniProt/SwissProt protein function and sequence, BindingDB compound--target measurements, Reactome pathway context, STRING interaction context, and LINCS L1000 perturbation signatures~\citep{uniprot,bindingdb,reactome,string,lincs_l1000}. Drug-anchored chains connect DrugBank/PubChem molecular identity to L1000 transcriptional responses through shared compound identifiers~\citep{drugbank,pubchem,lincs_l1000}. Cell-anchored chains connect CellxGene expression profiles to SwissProt annotations of marker genes~\citep{cellxgene,uniprot}. Additional chains use Hetionet, DRKG, STRING, and OmniPath/DoRothEA to link compounds, proteins, pathways, transcription factors, and cell-domain expression evidence~\citep{hetionet,drkg,string,dorothea}. These records are cross-domain because the supervision is the structured co-occurrence of molecular, protein, pathway, perturbation, and cell-context evidence, rather than a single-database fact or an unrelated mixture of domains.

\section{In-house Instruction Dataset Details}\label{C}

\paragraph{Protein binding.} We construct a protein-binding instruction dataset from structural complex records, not from free-text protein descriptions. For protein--ligand binding, we use PLINDER experimental and predicted-linked complexes~\citep{plinder}; for protein--protein binding, we use PINDER holo complexes~\citep{pinder}, PPIRef50K~\citep{ppiref}, DIPS-Plus~\citep{dipsplus}, PDB-derived domain-domain complexes~\citep{pdb}, and high-confidence AlphaFold-derived complexes~\citep{afdb}; and for protein--peptide binding, we use PDB direct complexes, Propedia~\citep{propedia}, and PPIKB-derived peptide--protein records~\citep{ppikb}. For each complex, we parse the target and binder chains, compute heavy-atom contacts using a 5.0 \AA{} cutoff, derive target-side and binder-side interface residues, and retain contact metadata such as contact count and contact type. These structured interfaces are then rendered into answer-supervised tasks: protein--ligand examples ask for binding-site residue listing, masked binding-site recovery, ligand-conditioned protein scaffolding, observed-interface summarization, or reference binder sequence generation; protein--protein examples ask for a binder sequence conditioned on a target protein and epitope residues; and protein--peptide examples ask for target-interface residues, peptide-interface residues, masked peptide-interface recovery, peptide binder generation, or protein--peptide interface summarization. We apply source-specific filters on sequence length, ligand size, contact count, interface size, and experimental resolution where available, deduplicate by unit/interface/prompt hashes, and exclude protein sequences overlapping binder-generation evaluation examples. The append step selects 12.0M binding instruction records before the final context-length filtering stage.

\paragraph{DNA/RNA feature localization.} DNA and RNA feature-localization instructions are constructed as exact span-recovery tasks over nucleotide sequences. For DNA, we use ENCODE SCREEN cCRE intervals and ENCODE open-chromatin peaks~\citep{encode}, GENCODE reference genome and transcript annotations for transcription start sites and splice donor/acceptor sites~\citep{gencode}, and JASPAR position-frequency matrices for transcription-factor motif hits~\citep{jaspar}. Each example extracts a sequence window containing exactly one target feature, surrounds it with flanking sequence, and asks the model to return valid JSON with the feature label, 1-indexed inclusive start and end coordinates, and the exact subsequence copied from the input. For RNA, we build the same instruction interface from GENCODE CDS/UTR annotations~\citep{gencode}, RNAcentral exon-junction records~\citep{rnacentral}, Rfam family hits~\citep{rfam}, miRBase mature-miRNA segments~\citep{mirbase}, and GtRNAdb tRNA anticodons~\citep{gtrnadb}. We reject ambiguous windows, windows with excessive unknown bases, features that do not occur exactly once when uniqueness is required, and records whose target subsequence fails a deterministic self-check. This makes regulatory and transcript annotations trainable as precise language-mediated extraction tasks, rather than as passive sequence narratives.

\begin{table*}[t]
\centering
\footnotesize
\caption{Full subtask performance for the small molecule domain. Bold indicates the best model in each scored row.}
\label{tab:appendix-small-molecule}

\setlength{\tabcolsep}{5pt}
\renewcommand{\arraystretch}{1.0}

\begin{tabularx}{0.7\textwidth}{
>{\raggedright\arraybackslash\bfseries}p{0.25\textwidth}
>{\raggedright\arraybackslash}X
r
r
}
\toprule
Subtask & \textbf{Metric} & \textbf{Base} & \textbf{Ours} \\
\midrule

\multirow{4}{*}{\makecell[l]{Description-guided\\molecular design}}
& Validity $\uparrow$ & 0.200 & \textbf{0.890} \\
& MACCS Tanimoto $\uparrow$ & 0.237 & \textbf{0.454} \\
& Morgan Tanimoto $\uparrow$ & 0.097 & \textbf{0.149} \\
& SMILES Levenshtein $\downarrow$ & 191.1 & \textbf{99.7} \\
\midrule

\multirow{4}{*}{Molecular reconstruction}
& Validity $\uparrow$ & 0.060 & \textbf{0.930} \\
& MACCS Tanimoto $\uparrow$ & 0.425 & \textbf{0.508} \\
& Morgan Tanimoto $\uparrow$ & 0.180 & \textbf{0.202} \\
& SMILES Levenshtein $\downarrow$ & 103.0 & \textbf{50.7} \\
\midrule

\multirow{3}{*}{Molecular recognition}
& Both fields exact $\uparrow$ & 0.220 & \textbf{0.290} \\
& result\_1 exact $\uparrow$ & 0.300 & \textbf{0.500} \\
& result\_2 exact $\uparrow$ & 0.320 & \textbf{0.380} \\
\midrule

\multirow{5}{*}{Forward synthesis}
& Canonical exact $\uparrow$ & 0.000 & \textbf{0.140} \\
& Validity $\uparrow$ & 0.340 & \textbf{1.000} \\
& MACCS Tanimoto $\uparrow$ & 0.334 & \textbf{0.756} \\
& Morgan Tanimoto $\uparrow$ & 0.177 & \textbf{0.578} \\
& SMILES Levenshtein $\downarrow$ & 64.9 & \textbf{16.2} \\
\bottomrule
\end{tabularx}
\end{table*}
\begin{table*}[t]
\centering
\footnotesize
\caption{Full subtask performance for the protein domain. Bold indicates the best model in each scored row.}
\label{tab:appendix-protein}

\setlength{\tabcolsep}{5pt}
\renewcommand{\arraystretch}{1.0}

\begin{tabularx}{0.7\textwidth}{
>{\raggedright\arraybackslash\bfseries}p{0.25\textwidth}
>{\raggedright\arraybackslash}X
r
r
}
\toprule
Subtask & \textbf{Metric} & \textbf{Base} & \textbf{Ours} \\
\midrule

\multirow{2}{*}{\makecell[l]{InterPro function\\prediction}}
& InterPro ID precision $\uparrow$ & 0.000 & \textbf{0.075} \\
& InterPro ID recall $\uparrow$ & 0.000 & \textbf{0.035} \\
\midrule

\multirow{4}{*}{\makecell[l]{Text-conditioned functional\\protein design}}
& Functional consistency $\uparrow$ & 0.321 & \textbf{0.427} \\
& Nondegeneracy $\uparrow$ & 0.050 & \textbf{0.930} \\
& pLDDT $\uparrow$ & 0.439 & \textbf{0.472} \\
& pTM $\uparrow$ & 0.160 & \textbf{0.258} \\
\midrule

\multirow{3}{*}{Binder design}
& ipTM $\uparrow$ & 0.131 & \textbf{0.289} \\
& Complex pLDDT $\uparrow$ & 0.452 & \textbf{0.646} \\
& Nondegeneracy $\uparrow$ & 0.120 & \textbf{1.000} \\
\bottomrule

\end{tabularx}
\end{table*}
\begin{table*}[t]
\centering
\footnotesize
\caption{Full subtask performance for the genomic sequence domain. Bold indicates the best model in each scored row.}
\label{tab:appendix-genomics}

\setlength{\tabcolsep}{5pt}
\renewcommand{\arraystretch}{1.0}

\begin{tabularx}{0.7\textwidth}{
>{\raggedright\arraybackslash\bfseries}p{0.25\textwidth}
>{\raggedright\arraybackslash}X
r
r
}
\toprule
Subtask & \textbf{Metric} & \textbf{Base} & \textbf{Ours} \\
\midrule

\multirow{4}{*}{\makecell[l]{cCRE class and\\genomic span}}
& Full JSON exact $\uparrow$ & 0.000 & \textbf{0.010} \\
& Coordinate exact $\uparrow$ & 0.000 & \textbf{0.420} \\
& Categorical exact $\uparrow$ & 0.030 & \textbf{0.290} \\
& Subsequence exact $\uparrow$ & 0.000 & \textbf{0.170} \\
\midrule

\multirow{4}{*}{\makecell[l]{Open chromatin\\peak and summit}}
& Full JSON exact $\uparrow$ & 0.000 & \textbf{0.080} \\
& Coordinate exact $\uparrow$ & 0.000 & \textbf{0.220} \\
& Categorical exact $\uparrow$ & 0.130 & \textbf{0.850} \\
& Subsequence exact $\uparrow$ & 0.000 & \textbf{0.150} \\
\midrule

\multirow{4}{*}{\makecell[l]{Splice-site type, strand,\\and motif span}}
& Full JSON exact $\uparrow$ & 0.130 & \textbf{1.000} \\
& Coordinate exact $\uparrow$ & 0.820 & \textbf{1.000} \\
& Categorical exact $\uparrow$ & 0.290 & \textbf{1.000} \\
& Subsequence exact $\uparrow$ & 0.210 & \textbf{1.000} \\
\midrule

\multirow{4}{*}{\makecell[l]{Rfam family and\\RNA element span}}
& Full JSON exact $\uparrow$ & 0.000 & \textbf{0.370} \\
& Coordinate exact $\uparrow$ & 0.790 & \textbf{0.840} \\
& Categorical exact $\uparrow$ & 0.800 & \textbf{0.840} \\
& Subsequence exact $\uparrow$ & 0.000 & \textbf{0.380} \\
\midrule

\multirow{4}{*}{\makecell[l]{tRNA amino acid and\\anticodon span}}
& Full JSON exact $\uparrow$ & 0.010 & \textbf{0.100} \\
& Coordinate exact $\uparrow$ & 0.270 & \textbf{0.400} \\
& Categorical exact $\uparrow$ & 0.010 & \textbf{0.120} \\
& Subsequence exact $\uparrow$ & 0.020 & \textbf{0.120} \\
\bottomrule

\end{tabularx}
\end{table*}
\begin{table*}[t]
\centering
\footnotesize
\caption{Full subtask performance for the cell/pathway domain. Bold indicates the best model in each scored row.}
\label{tab:appendix-cell-pathway}

\setlength{\tabcolsep}{5pt}
\renewcommand{\arraystretch}{1.0}

\begin{tabularx}{0.7\textwidth}{
>{\raggedright\arraybackslash\bfseries}p{0.25\textwidth}
>{\raggedright\arraybackslash}X
r
r
}
\toprule
Subtask & \textbf{Metric} & \textbf{Base} & \textbf{Ours} \\
\midrule

Cell type from marker genes
& Exact match $\uparrow$ & 0.470 & \textbf{0.580} \\
\midrule

Hallmark pathway from gene set
& Exact match $\uparrow$ & 0.520 & \textbf{0.750} \\
\midrule

\multirow{3}{*}{\makecell[l]{Gene regulation after\\K562 CRISPRi knockdown}}
& Direction accuracy $\uparrow$ & 0.000 & \textbf{0.468} \\
& Macro-F1 $\uparrow$ & 0.045 & \textbf{0.349} \\
& Candidate coverage $\uparrow$ & 0.000 & \textbf{0.675} \\
\bottomrule

\end{tabularx}
\end{table*}

\section{Full Experimental Results}\label{D}

In this section, we report subtask-wise performance across small molecules (\cref{tab:appendix-small-molecule}), proteins (\cref{tab:appendix-protein}), genomic sequences (\cref{tab:appendix-genomics}), and cell/pathway (\cref{tab:appendix-cell-pathway}).

\subsection{Small molecules}
For small molecules, the base model struggles to generate valid molecular outputs: generation validity never exceeds 0.20, and forward synthesis yields no exact-match products. In contrast, training on \system{} improves both validity and structural accuracy. The model trained on our corpus increases chemical validity to 0.89--1.00 and consistently achieves higher fingerprint similarity to the reference molecules. The most pronounced improvement is observed in forward synthesis, where correct products emerge despite zero exact matches from the vanilla model.

\subsection{Proteins}
In the protein domain, the vanilla model collapses to degenerate, repetitive sequences, with nondegeneracy near 0.05. We observe a paradoxical relationship between fold confidence (pLDDT) and nondegeneracy, suggesting that structural confidence alone can be misleading. We attribute this behavior to the tendency of structure prediction models to assign high confidence to low-complexity repetitive sequences when they match memorized local sequence patterns. For example, repeated motifs such as \texttt{QVQVQ...} may readily induce plausible local structural arrangements through learned subsequence priors despite lacking meaningful biochemical diversity. In contrast, our trained model raises nondegeneracy to 0.93 for monomer design and 1.00 for binder generation.

Additionally, we find that training on \system{} provides the smallest gains for protein function prediction. We read this as a coverage gap, where function-prediction instructions form a relatively small and narrow slice of the corpus next to the much larger generation and structure tasks; the model sees few examples to improve here. Scaling up and diversifying that supervision is the natural future direction.

\subsection{Genomic sequences}

Genomic sequence tasks are particularly challenging because they require recovering structured annotations rather than predicting a single label. The base model rarely recovers complete structured answers, with Full JSON exact-match scores remaining near zero across most tasks. In contrast, our trained model substantially improves genomic feature localization, achieving perfect performance on splice-site localization and large gains on cCRE, open-chromatin, and RNA-element localization. These improvements are particularly pronounced for structured span recovery; for example, Full JSON exact match on splice-site localization increases from 0.13 to 1.00, indicating that the model learns to identify and extract biologically meaningful sequence features.

\subsection{Cell/Pathway}

Cell and pathway tasks improve consistently across all subtasks. Cell-type recognition increases from 0.47 to 0.58, while hallmark-program recognition improves from 0.52 to 0.75. The largest gains are observed in perturbation-response prediction, where the vanilla model scores zero on both direction accuracy and candidate coverage, whereas our model reaches 0.468 direction accuracy and 0.675 candidate coverage. These results suggest that the corpus improves coverage across diverse cell-related tasks, whereas the vanilla model struggles on perturbation response (gene regulation) prediction.

\section{Evaluation Details}\label{E}
\begin{table}[t]
\centering
\footnotesize
\caption{Evaluation dataset statistics of \evalsuite{}.}
\label{tab:eval-statistics}
\setlength{\tabcolsep}{4pt}
\renewcommand{\arraystretch}{0.95}
\begin{tabularx}{.95\linewidth}{
  >{\raggedright\arraybackslash\bfseries}p{0.18\linewidth}
  >{\raggedright\arraybackslash}X
  >{\raggedleft\arraybackslash}p{0.10\linewidth}
}
\toprule
Domain & \textbf{Evaluation subtask} & \textbf{Queries} \\
\midrule

\multirow{4}{*}{Small molecules}
& Description-guided molecule design & 100 \\
& Molecular reconstruction & 100 \\
& Molecular recognition & 100 \\
& Forward synthesis prediction & 100 \\
\cmidrule(lr){2-3}
& \textbf{Subtotal} & \textbf{400} \\
\midrule

\multirow{3}{*}{Proteins}
& Text-conditioned protein design & 100 \\
& Binder generation & 100 \\
& Protein function prediction & 100 \\
\cmidrule(lr){2-3}
& \textbf{Subtotal} & \textbf{300} \\
\midrule

\multirow{5}{*}{Genomic sequences}
& cCRE localization & 100 \\
& Open-chromatin localization & 100 \\
& Splice-site localization & 100 \\
& Rfam hit localization & 100 \\
& tRNA anticodon localization & 100 \\
\cmidrule(lr){2-3}
& \textbf{Subtotal} & \textbf{500} \\
\midrule

\multirow{3}{*}{Cells/pathways}
& Tabula Sapiens cell-type classification & 100 \\
& Hallmark pathway classification & 100 \\
& Replogle K562 CRISPRi perturbation-response prediction & 100 \\
\cmidrule(lr){2-3}
& \textbf{Subtotal} & \textbf{300} \\
\midrule

\multirow{3}{*}{\makecell[l]{Cross-domain\\reasoning}}
& Protein function $\rightarrow$ pathway & 50 \\
& TF function $\rightarrow$ regulated target gene & 50 \\
& Small molecule $\rightarrow$ binding target, pathway & 50 \\
\cmidrule(lr){2-3}
& \textbf{Subtotal} & \textbf{150} \\
\midrule

\textbf{All domains}
& \textbf{Total queries}
& \textbf{1,650} \\
\bottomrule
\end{tabularx}
\end{table}

We evaluate \system{} with task-specific metrics that match the structure of each biological output. For all tasks, model generations are first parsed into the expected answer format and normalized with the same deterministic post-processing used for scoring. For the main summary table, scores are averaged first across the reported metrics within a subtask and then across subtasks within each biological domain. 
The statistics of our evaluation dataset is in \cref{tab:eval-statistics}, and the representative subtask-wise evaluation queries are shown in \cref{tab:eval-question-examples}. 
We detail the metrics we adopted for each domain as follows.

\textbf{Small molecules}

\begin{itemize}[topsep=0pt,itemsep=1mm, parsep=0pt, leftmargin=5mm]
    \item \textbf{Validity: }the fraction of model outputs that can be parsed as chemically valid SMILES by RDKit.
    \item \textbf{Exact match (exact): }compares the predicted and reference answers. For a canonical exact match, the predicted and reference molecules are matched after canonical SMILES normalization. 
    \item \textbf{MACCS and Morgan Tanimoto similarity: }fingerprint overlap between the predicted and reference molecules.
    \item \textbf{SMILES Levenshtein distance: }edit distance between canonical predicted and reference SMILES strings. 
\end{itemize}

\textbf{Proteins}

\begin{itemize}[topsep=0pt,itemsep=1mm, parsep=0pt, leftmargin=5mm]
    \item \textbf{InterPro precision: }the fraction of predicted functional annotations that are correct.
    \item \textbf{InterPro recall: }the fraction of gold annotations recovered by the model.
    \item \textbf{Functional consistency: }the cosine similarity between embeddings of the generated protein and the reference protein from ESM-C~\citep{esmfold2}. 
    \item \textbf{Nondegeneracy: }whether generated protein sequences avoid low-complexity or repetitive artifacts. A sequence can be syntactically valid while still being biologically implausible; nondegeneracy separates valid amino-acid output from collapsed or repetitive generations.
    \item \textbf{Fold quality: }pLDDT, pTM, and ipTM are fold-confidence metrics computed from ESMFold2-fast~\citep{esmfold2}. These metrics assess the foldability of the given sequence in either a single-chain or a complex context.
\end{itemize}

\textbf{Genomic sequences}

\begin{itemize}[topsep=0pt,itemsep=1mm, parsep=0pt, leftmargin=5mm]
    \item \textbf{Full JSON exact match (Full JSON exact): }whether the complete predicted JSON object matches the gold answer after canonical serialization.
    \item \textbf{Coordinate exact match (Coordinate exact): }compares only coordinate fields such as start and end positions. 
    \item \textbf{Categorical exact match (Categorical exact): }compares non-coordinate annotation fields, such as feature type, class, strand, family, or motif label, depending on the subtask. 
    \item \textbf{Subsequence exact match (Subsequence exact): }compares the sequence field copied from the input window. 
\end{itemize}

\textbf{Cell/Pathway}

\begin{itemize}[topsep=0pt,itemsep=1mm, parsep=0pt, leftmargin=5mm]
    \item \textbf{Direction accuracy: }the fraction of candidate genes whose predicted direction matches the gold differential-expression direction after perturbation.
    \item \textbf{Macro-F1: }computes F1 separately for up-regulated and down-regulated gene sets and averages the two classes.
    \item \textbf{Candidate coverage: }the fraction of candidate genes for which the model emits a valid direction prediction.
\end{itemize}

\textbf{Cross-domain understanding}
\begin{itemize}[topsep=0pt,itemsep=1mm, parsep=0pt, leftmargin=5mm]
    \item \textbf{Accuracy: }each task is a four-way multiple-choice question; we select the option with the highest model likelihood and count it correct when it matches the gold choice.
\end{itemize}

\begin{table*}[t]
\centering
\scriptsize
\setlength{\tabcolsep}{3.5pt}
\renewcommand{\arraystretch}{1.0}
\caption{Example evaluation questions by subtask. For convenience, each example is the final held-out question only, with few-shot demonstrations removed.}
\label{tab:eval-question-examples}
\begin{tabularx}{\textwidth}{p{2.1cm}p{2.8cm}X}
\toprule
\textbf{Domain} & \textbf{Evaluation subtask} & \textbf{Example question} \\
\midrule
Small molecule & Description-guided molecule design &
Synthesize a molecule that matches the given characteristics. The molecule appears as colorless crystals. Insoluble in water. \\
\midrule
Small molecule & Molecular reconstruction &
Reconstruct the molecule described below and answer with one SMILES string only. Begin with a six-membered benzene ring. At position 1, attach an amide connected to a piperidine ring whose nitrogen bears a sulfonyl-fused 1,2,5-oxadiazole/benzene system; at position 4, attach an isopropyl group. \\
\midrule
Small molecule & Molecular recognition &
Analyze the molecule and answer as \texttt{result\_1: <value>; result\_2: <value>}. SMILES: \texttt{<smiles>C(=CCC=CCC)CC\ldots</smiles>}. Task: ester. Report the number of esters and ester atom indices. \\
\midrule
Small molecule & Forward synthesis prediction &
Predict the product of the reaction from the given reactants and reagents: \texttt{<smiles>C1CCOC1.CCN(CC)\ldots</smiles>}. \\
\midrule
Protein & Text-conditioned protein design &
Create a protein sequence with the necessary features to perform the desired function. The designed protein must contain at least one GTP cyclohydrolase II domain, include an accessible Mg(2+) binding site, and contain a structurally stable nucleophile for GTP cyclohydrolase activity. \\
\midrule
Protein & Binder generation &
Design a de novo protein binder for a target protein. Use only the 20 standard amino acids and produce a single chain of exactly 204 residues. The binder should engage the listed target epitope residues on chain R; output only the binder sequence. \\
\midrule
Protein & Protein function prediction &
Given protein information for human SWI/SNF-related chromatin regulator SMARCA4, with InterPro annotations hidden, predict likely InterPro domain, family, repeat, site, or homologous-superfamily IDs. Output only semicolon-separated InterPro IDs. Sequence: \texttt{<protein>XRRDTALETALN\ldots</protein>}. \\
\midrule
Genomics & cCRE localization &
Given a human GRCh38 DNA sequence containing exactly one candidate cis-regulatory element, locate the cCRE and report its class and exact span. Return JSON with one region containing the label, start, and end. Sequence: \texttt{<dna>TCAAAAGAAAAA\ldots</dna>}. \\
\midrule
Genomics & Open-chromatin localization &
Given a human GRCh38 DNA sequence measured in K562, locate the single open-chromatin peak and its summit position. Return JSON with assay, biosample, start, end, and summit position. Sequence: \texttt{<dna>AGTGTGCTCACA\ldots</dna>}. \\
\midrule
Genomics & Splice-site localization &
Given a human GRCh38 DNA sequence containing one annotated exon-intron boundary, locate the splice site and report site type, strand, boundary position, and 2-bp motif span. Sequence: \texttt{<dna>CCTGGGTCAGAG\ldots</dna>}. \\
\midrule
Genomics & Rfam hit localization &
Given an RNA sequence containing one conserved non-coding RNA family element, identify the Rfam family and report the element span. Sequence: \texttt{<rna>GCCUGGCUGGCU\ldots</rna>}. \\
\midrule
Genomics & tRNA anticodon localization &
Given a tRNA sequence, locate the annotated anticodon triplet and report the carried amino acid. Sequence: \texttt{<rna>GCCGCAAUAGCU\ldots</rna>}. \\
\midrule
Cell/pathway & Cell type classification &
Given marker genes \texttt{CD24, PAX5, DNTT, PTPRC, \ldots} from the immune system, choose the single best supported cell type from candidates such as effector CD4+ T cells, memory CD4+ T cells, Pro-B cells, naive T cells, and classical monocytes. Return only the option letter. \\
\midrule
Cell/pathway & Hallmark pathway classification &
Given gene set \texttt{WNT1, CSNK1E, JAG1, DVL2, \ldots}, choose which biological program or pathway is most strongly represented from candidate Hallmark programs. Return only the option letter. \\
\midrule
Cell/pathway & Perturbation-response prediction &
Cell line: K562. Perturbation type: CRISPRi knockdown. Perturbed gene: \texttt{HINFP}. Given candidate genes \texttt{TNNI3, AIF1, MT-ND1, MT-CO3, \ldots}, predict which are up- or down-regulated and return JSON with \texttt{up} and \texttt{down} lists. \\
\midrule
Cross-domain & Protein function $\rightarrow$ pathway &
In which of the following pathways does this human protein participate? Protein description: ``Receptor for ghrelin, coupled to G-alpha-11 proteins''. 
\\
\midrule
Cross-domain & TF function $\rightarrow$ regulated gene &
A human transcription factor is described below and directly stimulates the transcription of one of the following genes (OmniPath). Which one? Function: ``Transcriptional activator which binds specifically to the MEF2 element\ldots found in numerous muscle-specific, growth factor- and stress-induced genes''. 
\\
\midrule
Cross-domain & Small molecule $\rightarrow$ target, pathway &
The small molecule below binds a human protein target (DRKG). Which option correctly pairs its target with a pathway that target participates in? SMILES: \texttt{<smiles>Cc1cccc\ldots</smiles>}. 
\\
\bottomrule
\end{tabularx}
\end{table*}

\section{Example Corpus Records}\label{F}

We provide representative records from \system{} for each major corpus-construction stage described in \cref{2_method}. These examples illustrate how structured biological resources are transformed into language-model-readable training records, from dataset refinement and feature enrichment to instruction-based supervision.

\subsection*{Dataset Refinement (\cref{2.2})}

The examples below illustrate how biological databases are converted into natural-language records while preserving the underlying biological information.

\textbf{Example 1. Protein--ligand binding record}
\begin{mdframed}
\fvset{
  breaklines=true,
  breakanywhere=true,
  breakautoindent=false,
  breaksymbolleft={},
  breaksymbolright={},
  breaksymbol={},
  breakindent=0pt,
  breakafter={},
  breakbefore={}
}
\begin{Verbatim}[fontsize=\scriptsize]
The ligand 2-{[4-(but-2-yn-1-ylamino)benzene]sulfonyl}ethane-1-thiol
(SMILES: <smiles>CC#CCNc1ccc(cc1)S(=O)(=O)CCS</smiles>,
InChI: <inchi>InChI=1S/C12H15NO2S2/c1-2-3-8-13-11-4-6-12(7-5-11)17(14,15)10-9-16/h4-7,13,
16H,8-10H2,1H3</inchi>)
was tested against the protein Disintegrin and metalloproteinase domain-containing protein 17 [215-477,S266A,N452Q].

The target protein name is Homo sapiens of ADA17_HUMAN and chain sequence is
<protein>RADPDPMKNTCKLLVVADHRFYRYMGRGEESTTTNYLIELIDRVDDIYRNTAWDNAGFKGYGIQIEQIRILKSPQEVKPGE
KHYNMAKSYPNEEKDAWDVKMLLEQFSFDIAEEASKVCLAHLFTYQDFDMGTLGLAYVGSPRANSHGGVCPKAYYSPVGKKNIYLNSGLT
STKNYGKTILTKEADLVTTHELGHNFGAEHDPDGLAECAPNEDQGGKYVMYPIAVSGDHENNKMFSQCSKQSIYKTIESKAQECFQERSN
KV</protein>.
PDB IDs of target chain are 2A8H,1ZXC,3G42,2I47,2OI0,3EWJ,3EDZ,3E8R,3B92,3CKI,1BKC

The experiment was performed at pH 7.5000 and 22.00 C.
The ligand exhibited strong binding strength, with a reported Ki (nM) of 2.
\end{Verbatim}
\end{mdframed}

\textbf{Example 2. Cellular perturbation record}
\begin{mdframed}
\fvset{
  breaklines=true,
  breakanywhere=true,
  breakautoindent=false,
  breaksymbolleft={},
  breaksymbolright={},
  breaksymbol={},
  breakindent=0pt,
  breakafter={},
  breakbefore={}
}
\begin{Verbatim}[fontsize=\scriptsize]
In HA1E (kidney) cells, palmitoylethanolamide (<smiles>CCCCCCCCCCCCCCCC(=O)NCCO</smiles>) at 0.04 uM for 24 h up-regulates SUV39H1, CLPX, SRC and down-regulates GJA1, HMGCS1, DNM1L (|z| >= 2.0), based on LINCS L1000 signature REP.A023_HA1E_24H:B24.
\end{Verbatim}
\end{mdframed}

\subsection*{Tool-Computed Feature Narratives (\cref{2.3})}

These examples demonstrate records enriched with tool-computed biological features that are not explicitly present in the original source databases.

\textbf{Example 3. Molecular feature narrative}
\begin{mdframed}
\fvset{
  breaklines=true,
  breakanywhere=true,
  breakautoindent=false,
  breaksymbolleft={},
  breaksymbolright={},
  breaksymbol={},
  breakindent=0pt,
  breakafter={},
  breakbefore={}
}
\begin{Verbatim}[fontsize=\scriptsize]
The molecule has the SMILES representation: <smiles>C1C=CC2=CSC=C2O1</smiles>
and its IUPAC systematic name is 2H-thieno[3,4-b]pyran. It also has the corresponding InChI: <inchi>InChI=1S/C7H6OS/c1-2-6-4-9-5-7(6)8-3-1/h1-2,4-5H,3H2</inchi>.

The compound contains 15 atoms, including 9 heavy atoms, and 16 bonds. The molecule is Achiral, and it has 0 charged atoms, and a total formal charge of 0. Its molecular formula is C7H6OS, giving an exact mass of 138.01393598 g/mol and a molecular weight of 138.19 g/mol.

The molecule features 2 hydrogen bond acceptors, 0 hydrogen bond donor, 0 rotatable bond, and an XlogP3-AA value of 1.9, indicating moderate polarity, balanced aqueous solubility and membrane permeability.

There are 2 rings in the molecular graph, including 1 aromatic, 1 aliphatic, and 0 saturated rings. Ring subtype counts include 0 aromatic carbocyclic, 1 aromatic heterocyclic, 0 aliphatic carbocyclic, 1 aliphatic heterocyclic, 0 saturated carbocyclic, and 0 saturated heterocyclic rings. The graph contains 2 heteroatoms, 0 amide bonds, 0 spiro atoms, and 0 bridgehead atoms. Its fraction sp3 is 0.14. No potential stereocenters are detected in the parsed molecular graph. The Murcko scaffold is C1=Cc2cscc2OC1. Graph complexity descriptors include Bertz complexity 236.8, Labute ASA 57.6 A^2, Hall-Kier alpha -0.8, and Kier kappa values 5.1, 1.9, and 0.8. The parsed graph has 46 valence electrons and 0 radical electrons.

Additional physicochemical descriptors: heavy-atom molecular weight 132.14, topological polar surface area (TPSA) 9.2 A^2, molar refractivity 38.8, QED drug-likeness 0.533, 0 Lipinski rule-of-five violations, NHOH count 0, N+O count 1. 

Structural alerts: 0 PAINS, 0 Brenk. Substructure map (0-based atom indices follow the given SMILES): Functional groups -> none. Aromatic rings -> 1 thiophene at [3, 4, 5, 6, 7]. Ring-junction atoms (shared by two rings): [3, 7]. Per-atom detail: Atom 5 is S (aromatic); 1-hop neighbors: 4 (aromatic), 6 (aromatic); 2-hop neighbors [3, 7]; 3-hop neighbors [2, 8]. Atom 0 is C; 1-hop neighbors: 1 (single), 8 (single); 2-hop neighbors [2, 7]; 3-hop neighbors [3, 6].
\end{Verbatim}
\end{mdframed}

\textbf{Example 4. Spatial transcriptomics narrative}
\begin{mdframed}
\fvset{
  breaklines=true,
  breakanywhere=true,
  breakautoindent=false,
  breaksymbolleft={},
  breaksymbolright={},
  breaksymbol={},
  breakindent=0pt,
  breakafter={},
  breakbefore={}
}
\begin{Verbatim}[fontsize=\scriptsize]
HuBMAP tissue-context spatial record:

sample=HBM332.KQWB.454,
assay=Slide-seq [Salmon],
organism=Homo sapiens,
region=Kidney (Right),
spot=CAGATCCACAACGC,
coordinate=x=1977.5, y=4192.2,
annotation=not reported.

Observed RNA capture:
total_counts=222.1,
detected_genes=236.

Retained spot-local transcripts:
MALAT1, MT-RNR2, CYTB, ND2, UMOD, ALDOB, ATP6, GUK1.

Spatial context:
12 neighboring spots within 50 um,
nearest_neighbor=17.4 um.
\end{Verbatim}
\end{mdframed}

\subsection*{Instruction-Tuning Datasets (\cref{2.4})}

The examples below illustrate instruction--answer pairs designed to capture biological tasks that are difficult to obtain from conventional text-based biological datasets.

\textbf{Example 5. Masked protein binding-site infilling}
\begin{mdframed}
\fvset{
  breaklines=true,
  breakanywhere=true,
  breakautoindent=false,
  breaksymbolleft={},
  breaksymbolright={},
  breaksymbol={},
  breakindent=0pt,
  breakafter={},
  breakbefore={}
}
\begin{Verbatim}[fontsize=\scriptsize]
Binding-site residues in the protein sequence are masked for ligand-conditioned infilling. Predict the original amino acid identity at each masked binding-site position. Positions are 1-indexed within the protein chain.

Ligand (SMILES): <smiles>NCC1=CC([N+](=O)[O-])C=CC1=O</smiles>

Protein (chain A), binding-site residues masked:
<protein>MNTV<mask><mask><mask>SAPIEVTIGIDQYSFNVKENQPFHGIKDIPIGHVHVIHFQHADNSSMRYGYWFD
CRMGNF<mask>IQYDPKDGLYKMME<mask>RDGA<mask<mask>EN<mask>VHNFKERQMMVSYPKIDEDDTWYNLTEFVQ
MDKIRKIVRKDENQFSYVDSSMTTVQENELSDPAHSLNYTVINFKSREAIRPGHEMEDFLDKSYYLNTVMLQGIFKNSSNYFGEL
QFAFLNAMFFGNYGSSLQWHAMIELICSSATVPKHMLDKLDEILYYQIKTLPEQYSDILLNERVWNICLYSSFQKNSLHNTEKIM
ENKYPELL</protein>
Masked positions: A9, A10, A11, A72, A87, A92, A93, A96

List the recovered amino acids by position as chain:residue amino acid and 1-indexed position tokens (e.g., A:V15).

Answer: A:P9, A:F10, A:T11, A:Y72, A:E87, A:K92, A:F93, A:I96
\end{Verbatim}
\end{mdframed}

\textbf{Example 6. Genomic motif localization}
\begin{mdframed}
\fvset{
  breaklines=true,
  breakanywhere=true,
  breakautoindent=false,
  breaksymbolleft={},
  breaksymbolright={},
  breaksymbol={},
  breakindent=0pt,
  breakafter={},
  breakbefore={}
}
\begin{Verbatim}[fontsize=\scriptsize]
You are given a human GRCh38 genomic DNA sequence, taken from the genome forward (+) strand, containing one high-confidence JASPAR transcription factor (TF) binding motif plus flanking DNA. The target motif is a high-confidence JASPAR pattern embedded in flanking genomic DNA.

Sequence:
<dna>AATTCATTACAGGCTTCAGAGGCTGAGCTTTTTGTGTTTCATTGGTTGAACACTTGATTATGGATTACAACTTATTTTGAACCACA
AGCCATCAAAATTTAGATGACTTGTTTCCAAAACTGGAGTACTTTTTTTGTAGAGCATTTGGATTGACTCATGAAGTATGTTTTAAAGTTA
AAAGCTGCCTAATGTCAGCAATTTCCTATCATTTCATTATTTCAAGTGTCAAATATTTCTCAAATCCAACTTGACAGGCAGCAAAGAGAGA
ACCGCACACTCTACTGATGAAATGTTTTTAATCCAATTTA</dna>

Task: Locate the TF motif hit.

- "strand" is "+" if the motif matches the sequence as written,
  or "-" if it matches the reverse complement.
- "subsequence" is always the forward-strand characters from
  start to end, regardless of "strand".

Coordinate rules:
- Positions are 1-indexed.
- All reported interval endpoints are inclusive.
- The output subsequence field must exactly match the characters
  of the input sequence over the interval specified for that task.
- The sequence contains exactly ONE target element.
  Return exactly one object.

Output rules:
- Output valid JSON only. No markdown fences, no comments,
  no explanation.
- Use exactly the keys shown. Do not add or rename keys.

Return JSON only:

{"motif_hits":[{"motif_name":"<string>","start":<int>,"end":<int>,
"strand":"<+ or ->","subsequence":"<string>"}]}

Answer:
{"motif_hits":[{"motif_name":"FOS","start":151,"end":158,
"strand":"+","subsequence":"TGACTCAT"}]}
\end{Verbatim}
\end{mdframed}

\end{CJK*}
\end{document}